\documentclass{article}
\usepackage[utf8]{inputenc}

\usepackage[a4paper,top=3cm,bottom=2cm,left=3cm,right=3cm,marginparwidth=1.75cm]{geometry}

\usepackage[toc,page]{appendix}

\usepackage{amsmath,amssymb}
\usepackage{graphicx}
\usepackage[colorinlistoftodos]{todonotes}
\usepackage[colorlinks=true, allcolors=black]{hyperref}
\usepackage{amsthm, mathtools,amssymb}
\usepackage{graphicx,color}
\usepackage{tabularx}
\usepackage{tikz}
\usepackage{pgfplots}
\usepackage{pst-plot}
\usepackage{pstricks}
\usepackage{tabto}
\usepackage{braket}
\usepackage{enumitem}
\usepackage{soul}
\usepgfplotslibrary{fillbetween}
\usetikzlibrary{patterns}
\usetikzlibrary{shadings}
\usetikzlibrary{arrows}
\usetikzlibrary{calc}
\usetikzlibrary{decorations.markings}
\DeclareGraphicsExtensions{.pdf,.eps,.fig,.pdf_tex}
\usepackage{caption,subcaption,float}

\theoremstyle{plain} \numberwithin{equation}{section}
\newtheorem{theorem}{Theorem}[section]
\newtheorem{corollary}[theorem]{Corollary}

\newtheorem{lemma}[theorem]{Lemma}
\newtheorem{proposition}[theorem]{Proposition}
\theoremstyle{definition}
\newtheorem{definition}[theorem]{Definition}

\newtheorem{remark}[theorem]{Remark}

\title{Gluing of Graph Laplacians and Their Spectra}
\date{}

\author{
 Ivan Contreras
\and Michael Toriyama 
\and Chengzheng Yu
}

\begin{document}
\maketitle


\begin{abstract} \label{Abstract_Section}
We study two different types of gluing for graphs: interface (obtained by choosing a common subgraph as the gluing component) and bridge gluing (obtained by adding a set of edges to the given subgraphs).
We introduce formulae for computing even and odd Laplacians of graphs obtained by gluing, as well as their spectra. We subsequently discuss applications to quantum mechanics and bounds for the Fiedler value of the gluing of graphs.
\end{abstract}


\section{Introduction} \label{Introduction_Section}
Spectral graph theory plays a fundamental role in several areas of research, including network theory, data analysis, and modeling physical processes. The main characters involved in spectral graph analysis are the adjacency, incidence, and Laplacian matrices. The \emph{adjacency matrix} $A_{\Gamma}$ of a finite graph $\Gamma=(V,E)$ is defined by
\begin{equation*}
A_{\Gamma}:=A(i,j) = \left\{\begin{matrix}
1 & \text{if } v_i \text{ is adjacent to } v_j\\ 
0 & \text{otherwise.}
\end{matrix}\right.
\end{equation*}
The \emph{incidence matrix} $\mathcal{I}_{\Gamma}$ of a finite oriented graph $\Gamma$ is a $|V|\times |E|$-matrix given by \begin{equation*}
\mathcal{I}_{\Gamma}:=\mathcal{I}(i,j)=\left\{\begin{matrix}
1 & \text{if } e_{j} \text{ ends at } v_{i}\\ 
-1 & \text{if } e_{j} \text{ starts at } v_{i}\\ 
0 & \text{otherwise.}
\end{matrix}\right.
\end{equation*}

The even Laplacian matrix, denoted by $\Delta^+_{\Gamma}$, is a discrete analog of the Laplace operator in $\mathbb R^n$ and provides a measurement of how a function on $\Gamma$ differs at one vertex from its values at neighboring vertices. This definition suggests that there is another self-adjoint discrete operator, which measures the difference of ``edge-values" of a function on $\Gamma$, relative to the neighbors. We will refer to this operator as the \emph{odd graph Laplacian} $\Delta^-_{\Gamma}$ (see e.g. \cite{2017_Yu_Hyperwalk, 2017_Yu_Superwalk}).

A question arising in spectral graph theory is how the two Laplacian matrices and their spectra evolve when two initially disjoint graphs are glued in various ways. In this manuscript, we introduce two types of gluing called \emph{interface gluing} and \emph{bridge gluing}. We study the graph Laplacians and their spectra after gluing, and we suggest some applications of these results in spectral graph theory, network analysis, and quantum mechanics. In quantum mechanics particularly, we motivate a discretized version of the Schr\"{o}dinger equation inspired by the locality principle.

Additionally, we explore the implications of interface and bridge gluing on the Fiedler value $F_{\Gamma}$ \cite{2007_Abreu}, also known as algebraic connectivity of graphs. The value of $F_{\Gamma}$ is defined as the smallest nonzero eigenvalue of $\Delta^{+}_{\Gamma}$. This is always non-negative because $\Delta^{+}_{\Gamma}$ is positive semidefinite, and it provides a measurement of how well $\Gamma$ is connected. Bounds for the Fiedler value have been developed for large planar graphs \cite{2013_Barriere} and trees \cite{2007_Yuan}, however explicit formulas to calculate this quantity are computationally hard to achieve. We therefore propose an algorithm for computing this quantity when two graphs are glued via interfaces and bridges. 

This paper is organized as follows: we explicitly define in Section~\ref{Definitions_Section} the notions of interface and bridge gluing. We then explore how the even and odd Laplacian matrices, as well as their spectra depend on those of the original graphs for interface gluing (Section~\ref{InterfaceGluing_Section}) and bridge gluing (Section~\ref{BridgeGluing_Section}). Section~\ref{Examples_Section} is dedicated to providing examples of our main results, and we propose some applications in network analysis and graph quantum mechanics, following Witten's approach to supersymmetry via Morse theory \cite{1982_Witten} (Section~\ref{Conclusion_Section}).


\section*{Acknowledgements}
This work was supported by the National Science Foundation under Grant Number DMS-1449269. Any opinions, findings, and conclusions or recommendations expressed in this material are those of the author(s) and do not necessarily reflect the views of the National Science Foundation. This work was supported by the Illinois Geometry Lab (IGL) in the Department of Mathematics at University of Illinois at Urbana-Champaign. M.T. acknowledges the Illinois Scholars Undergraduate Research (ISUR) program and funding provided by Intel and Semiconductor Research Corporation. I.C. thanks Andrew Eberlein, Pavel Mnev, Hadrian Quan, and Yunting Zhang for useful discussions and comments.


\section*{Notation and Conventions}

All the graphs $\Gamma$ considered througout the paper are finite. We denote the vertex and edge sets of $\Gamma$ by $V(\Gamma)$ and $E(\Gamma)$, respectively. We denote by $M(i,j)$ the $(i,j)$-index of the matrix $M$ except in Sections~\ref{InterfaceGluingSpectra_Subsection} and~\ref{BridgeGluingSpectra_Subsection}, where we denote the $(i,j)$-index of matrix $M$ by $M^{(i,j)}$. We also denote the $(i,j)$-minor of matrix $M$ by $M_{(i,j)}$. The characteristic polynomial of matrix $M$ will be denoted by $p_{M}(\lambda)$.


\section{Definitions and General Properties of the Laplacian} \label{Definitions_Section}

In this section, we state the main definitions and properties that will be needed for the subsequent sections. 

\begin{definition} \label{EvenGraphLaplacian_Definition}
The \emph{even graph Laplacian matrix} $\Delta_{\Gamma}^{+}$ of a graph $\Gamma$ is a $|V(\Gamma)| \times |V(\Gamma)|$-matrix defined by
\begin{equation*}
 \Delta_{\Gamma}^{+}(i,j) = \begin{cases}
\text{val}(v_i) & \text{if } i=j \\
-1 & \text{if } v_i \text{ is adjacent to } v_j \\
0 & \text{otherwise.}
\end{cases}
\end{equation*}
\end{definition}

It is not hard to prove that with this definition, we also obtain that
\begin{equation*}
\Delta_{\Gamma}^{+} = \mathcal{I}\mathcal{I}^{t},
\end{equation*}
where $\mathcal{I}$ is the incidence matrix of $\Gamma$. Observe that $\Delta_{\Gamma}^{+}$ is independent of the orientation of $\Gamma$.

\begin{definition} \label{OddGraphLaplacian_Definition}
The \emph{odd graph Laplacian matrix} $\Delta_{\Gamma}^{-}$ of a directed graph $\Gamma$ is a $|E(\Gamma)| \times |E(\Gamma)|$-matrix defined by
\begin{equation*}
\Delta_{\Gamma}^{-} := \mathcal{I}^{t} \mathcal{I},
\end{equation*}
\end{definition}

which depends on the orientation of $\Gamma$. Later, we will see the relationship between $\Delta_{\Gamma}^{-}$, $\Delta_{\Gamma}^{+}$, and the topology of $\Gamma$. 

\begin{definition} \label{CharacteristicPolynomialRatio_Definition}
We denote by $\mathcal{E}(\Gamma)$ the ratio of the characteristic polynomials of the even and odd Laplacians of $\Gamma$, i.e. $\mathcal{E}(\Gamma) := \frac{p_{\Delta^+_{\Gamma}}(\lambda)}{p_{\Delta^-_{\Gamma}}(\lambda)}$.
\end{definition}

The following definitions describe the gluing procedure for graphs. Definitions \ref{Interface_Definition} and \ref{GluingInterface_Definition} lay the groundwork for interface gluing.

\begin{definition}\label{Interface_Definition}
Let $\Gamma_1$ and $\Gamma_2$ be two oriented graphs. If $\Gamma_1^{\partial}$ and $\Gamma_2^{\partial}$ are two isomorphic directed subgraphs of $\Gamma_1$ and $\Gamma_2$ respectively, then $I = \Gamma_1 \sqcap \Gamma_2 = \Gamma_1^{\partial} = \Gamma_2^{\partial}$ is an \emph{interface} of $\Gamma_1$ and $\Gamma_2$.
\end{definition}

\begin{definition}\label{GluingInterface_Definition}
Let $\Gamma_1$ and $\Gamma_2$ be two oriented graphs with interface $I = \Gamma_1 \sqcap \Gamma_2$. Then, the \emph{interface gluing} of the two graphs $\Gamma_1 \sqcup_I \Gamma_2$ is defined by
\begin{equation*}
V(\Gamma_1 \sqcup_I \Gamma_2)=\big( V(\Gamma_1) \setminus V(I) \big) \cup \big( V(\Gamma_2) \setminus V(I) \big) \cup V(I)
\end{equation*}
and
\begin{equation*}
E(\Gamma_1 \sqcup_I \Gamma_2)=\big( E(\Gamma_1) \setminus E(I) \big) \cup \big( E(\Gamma_2) \setminus E(I) \big) \cup E(I).
\end{equation*}
\end{definition}

On the other hand, Definitions \ref{Bridge_Definition} and \ref{BridgeGluing_Definition} define the concept of bridge gluing of two graphs.

\begin{definition}\label{Bridge_Definition}
Let $\Gamma_1$ and $\Gamma_2$ be two oriented graphs, and $\{v^1_1, \ldots, v^1_k\} \subseteq V(\Gamma_1)$, $\{v^2_1, \ldots, v^2_k\} \subseteq V(\Gamma_2)$. If we use $k$ edges $\{e_1, \ldots, e_k\}$ to connect $k$ pairs of distinct vertices $(v^1_1, v^2_1), \ldots, (v^1_k, v^2_k)$, then a \emph{bridge graph} $B$ between $\Gamma_1$ and $\Gamma_2$ is a graph where $V(B) = \{v^1_1, \ldots, v^1_k, v^2_1, \ldots, v^2_k\}$ and $E(B) = \{e_1, \ldots, e_k\}$.
\end{definition}

\begin{definition} \label{BridgeGluing_Definition}
Let $\Gamma_1$ and $\Gamma_2$ be two graphs and $B$ a bridge graph. Then, the \emph{bridge gluing} of the two graphs $\Gamma_1 \sqcup_B \Gamma_2$ is defined by
\begin{equation*}
V(\Gamma_1 \sqcup_B \Gamma_2) = V(\Gamma_1) \cup V(\Gamma_2)
\end{equation*}
and
\begin{equation*}
E(\Gamma_1 \sqcup_B \Gamma_2) = E(\Gamma_1) \cup E(\Gamma_2) \cup E(B).
\end{equation*}
\end{definition}

\begin{remark}\label{ZetaConditionRemark}
Let $\Gamma$ be a oriented graph, and $e_i$, $e_j$ be two edges in $\Gamma$. Then, exactly one of the following conditions holds:

\begin{enumerate}
\item Two edges $e_i$ and $e_j$ are not incident ($e_i \cap e_j = \emptyset$);
\item Two edges $e_i$ and $e_j$ are incident, and they both start or end at the common vertex ($e_i \cap e_j = \{v\}$, and both $e_i$ and $e_j$ start or end at $v$);
\item Two edges $e_i$ and $e_j$ are incident, and one of them starts at the common vertex, while the other one ends at the common vertex ($e_i \cap e_j = \{v\}$, and $e_i$ starts at $v$, while $e_j$ ends at $v$).
\end{enumerate}

\end{remark}

Remark~\ref{ZetaConditionRemark} gives us all three cases between two edges in an oriented graph. In this paper, we will refer to them as the $\zeta$-conditions.


\subsection{Laplacians when Changing Orientations} \label{LaplacianChangeOrientation_Subsection}
As we remarked above, the even Laplacian is independent of the orientation (since it only depends on the valence and the adjacency of vertices), whereas the odd Laplacian is sensitive to changes in the orientation of edges. The following theorems give a precise dependency of $\Delta^{-}_{\Gamma}$ on the orientation.

\begin{proposition}\label{odd_l_prop}
Let $\Gamma = \big(V(\Gamma),E(\Gamma)\big)$ and $\Gamma_{\mathcal O} = \big(V(\Gamma_{\mathcal O}),E(\Gamma_{\mathcal O})\big)$ be two oriented graphs, such that $\Gamma$ and $\Gamma_{\mathcal O}$ are isomorphic as non-oriented graphs. Let $\mathcal O = \{p, q, \ldots, r\}$ be a subset of $\{1,2,\ldots,|E(\Gamma)|\}$ (and, equivalently, a subset of $\{1,2,\ldots,|E(\Gamma_{\mathcal O})|\}$) such that whenever $k \in \mathcal O$, $e_k$ has opposite orientations in $\Gamma$ and $\Gamma_{\mathcal O}$, and whenever $k \notin \mathcal O$, $e_k$ has the same orientation in $\Gamma$ and $\Gamma_{\mathcal O}$. Then we have
$$\Delta^-_{\Gamma_{\mathcal O}}(i,j)=\left\{\begin{matrix}
\Delta^-_{\Gamma}(i,j) & \text{if } i \in \mathcal O \text{ and } j \in \mathcal O\\ 
\Delta^-_{\Gamma}(i,j) & \text{if } i \notin \mathcal O \text{ and } j \notin \mathcal O\\ 
-\Delta^-_{\Gamma}(i,j) & \text{if } i \in \mathcal O \text{ and } j \notin \mathcal O\\
-\Delta^-_{\Gamma}(i,j) & \text{if } i \notin \mathcal O \text{ and } j \in \mathcal O.
\end{matrix}\right.$$
\end{proposition}

\begin{proof}
Since $\Delta^-=\mathcal{I}^t \mathcal{I}$, then it is a direct observation that $$\Delta^-_{\Gamma_{\mathcal O}}(i,j)=\left\{\begin{matrix}
(-e_i)^t(-e_j) = \Delta^-_{\Gamma}(i,j) & \text{if } i \in \mathcal O \text{ and } j \in \mathcal O\\ 
(e_i)^t(e_j) = \Delta^-_{\Gamma}(i,j) & \text{if } i \notin \mathcal O \text{ and } j \notin \mathcal O\\ 
(-e_i)^t(e_j) = -\Delta^-_{\Gamma}(i,j) & \text{if } i \in \mathcal O \text{ and } j \notin \mathcal O\\
(e_i)^t(-e_j) = -\Delta^-_{\Gamma}(i,j) & \text{if } i \notin \mathcal O \text{ and } j \in \mathcal O.
\end{matrix}\right.$$
\end{proof}

Proposition~\ref{odd_l_prop} gives an explicit relationship between $\Delta^-_{\Gamma}$ and $\Delta^-_{\Gamma_{\mathcal O}}$, therefore we can also provide a relationship between their eigenvalues.

\begin{theorem}\label{same_spectrum}
If $\Gamma$ and $\Gamma_{\mathcal O}$ are two oriented graphs which differ by the orientation of one or more edges, then $\Delta_\Gamma^+$, $\Delta_{\Gamma_{\mathcal O}}^+$, $\Delta_\Gamma^-$, and $\Delta_{\Gamma_{\mathcal O}}^-$ have the same spectrum.
\end{theorem}

\begin{proof}
By Lemma~\ref{Eigenspace_Lemma}, we have $\Delta^{+}_{\Gamma} = \mathcal{I}\mathcal{I}^t$ and $\Delta^{-}_{\Gamma} = \mathcal{I}^t\mathcal{I}$ share the same eigenvalues, and $\Delta^{+}_{\Gamma}$ does not depend on the orientation of the graph. Hence, $\Delta_\Gamma^+$, $\Delta_{\Gamma_{\mathcal O}}^+$, $\Delta_\Gamma^-$, and $\Delta_{\Gamma_{\mathcal O}}^-$ have the same spectrum.
\end{proof}


\subsection{Spectra of Even and Odd Laplacians} \label{SpectrumGluing_Section}

We have proven that the even and odd Laplacians of any graph are isospectral. The following proposition compares the multiplicity of the zero eigenvalue for both Laplacians.

\begin{proposition} \label{EvenOddCospectral_Proposition}
Suppose $\Gamma$ is a graph with $|E(\Gamma)| > 0$. If $\lambda \neq 0$ is an eigenvalue of $\Delta^{+}_{\Gamma}$ with multiplicity $m$, then $\lambda$ is also an eigenvalue of $\Delta^{-}_{\Gamma}$ with multiplicity $m$. Moreover, $\lambda = 0$ has multiplicity $b_0 - \left( |V(\Gamma)| - |E(\Gamma)| \right)$ for $\Delta^{-}_{\Gamma}$ where $b_0$ is the number of connected components of $\Gamma$.
\end{proposition}

\begin{proof}
By Lemma~\ref{Eigenspace_Lemma}, since $\Delta^{+}_{\Gamma} = \mathcal{I}\mathcal{I}^t$ and $\Delta^{-}_{\Gamma} = \mathcal{I}^t \mathcal{I}$, they share the same nonzero eigenvalues, each with the same multiplicity. Furthermore,
\begin{equation*}
b_0 - b_1 = \chi(\Gamma) = |V(\Gamma)| - |E(\Gamma)|,
\end{equation*} 
where $b_0$ and $b_1$ are the zeroth and first Betti numbers and $\chi(\Gamma)$ is the Euler characteristic. Since $b_1$ is the multiplicity of $\lambda = 0$ and $b_0$ is the number of connected components of $\Gamma$, as we wanted.
\end{proof}

\begin{remark}
In Sections~\ref{InterfaceGluingSpectra_Subsection} and~\ref{BridgeGluingSpectra_Subsection}, we will only consider gluing the spectra of even Laplacian matrices. The analogous gluing formulae for the odd Laplacian spectra follows from Proposition~\ref{EvenOddCospectral_Proposition}.
\end{remark}

\subsection{The Graph Laplacian and the Topology of Graphs}
As we will elaborate throughout the paper, there is a connection between the spectrum of the graph Laplacians and the topology of the corresponding graph $\Gamma$, following a discrete version of Hodge theory on manifolds.
\begin{proposition} \label{EulerCharacteristic_Proposition}
$\mathcal{E}(\Gamma) = (-\lambda)^{\chi(\Gamma)}$, where $\chi(\Gamma)$ is the Euler characteristic of $\Gamma$.
\end{proposition}

\begin{proof}
By Lemma~\ref{Eigenspace_Lemma}, if $\lambda$ is an eigenvalue of $\Delta^+_\Gamma$, then $\lambda$ is also an eigenvalue of $\Delta^-_\Gamma$. Moreover, if $\lambda \neq 0$, it is an eigenvalue in both $\Delta^+_\Gamma$ and $\Delta^-_\Gamma$ with the same multiplicity. Thus,
\begin{equation*}
\mathcal{E}(\Gamma) = \frac{p_{\Delta^+_\Gamma}(\lambda)}{p_{\Delta^-_\Gamma}(\lambda)} = \frac{(-\lambda)^\alpha(\lambda_1-\lambda)^{m_1}(\lambda_2-\lambda)^{m_2}\ldots(\lambda_k-\lambda)^{m_k}}{(-\lambda)^\beta(\lambda_1-\lambda)^{m_1}(\lambda_2-\lambda)^{m_2}\ldots(\lambda_k-\lambda)^{m_k}} = \frac{(-\lambda)^\alpha}{(-\lambda)^\beta} = (-\lambda)^{\alpha - \beta},
\end{equation*}
where $\alpha = \text{dim}(N(\Delta^+))$ and $\beta = \text{dim}(N(\Delta^-))$. Recall that in a graph $\Gamma$, the zeroth Betti number is $b_0 = \text{dim}(N(\Delta^+))$, and the first Betti number is $b_1 = \text{dim}(N(\Delta^-))$. Hence we have \begin{equation*}
\mathcal{E}(\Gamma) = (-\lambda)^{b_0 - b_1} = (-\lambda)^{\chi(\Gamma)},
\end{equation*}
where $\chi(\Gamma)$ is the Euler characteristic of $\Gamma$ and $\chi(\Gamma) = |V(\Gamma)| - |E(\Gamma)|$.
\end{proof}



\section{Interface Gluing}\label{InterfaceGluing_Section}

Some implications of interface gluing on the even and odd Laplacian matrices as well as the spectrum are presented in this section. The following is a result which relates the interface gluing of graphs and the Euler characteristic.

\begin{theorem} \label{InterfaceGluingRatio_Theorem}
Let $\Gamma_1$ and $\Gamma_2$ be two graphs with interface $I$. Then, it follows that $\mathcal{E}(\Gamma_1 \sqcup_I \Gamma_2) = \frac{\mathcal{E}(\Gamma_1)\mathcal{E}(\Gamma_2)}{\mathcal{E}(I)}$.
\end{theorem}

\begin{proof}
We have 
\begin{equation*}
\begin{split}
\chi(\Gamma_1 \sqcup_I \Gamma_2) & = |V(\Gamma_1 \sqcup_I \Gamma_2)| - |E(\Gamma_1 \sqcup_I \Gamma_2)| \\
& = \left( |V(\Gamma_1)| + |V(\Gamma_2)| - |V(I)| \right) - \left( |E(\Gamma_1)| + |E(\Gamma_2)| - |E(I)| \right) \\
& = \left( |V(\Gamma_1)| - |E(\Gamma_1)| \right) + \left( |V(\Gamma_2)| - |E(\Gamma_2)| \right) - \left( |V(I)| - |E(I)| \right) \\
& = \chi(\Gamma_1) + \chi(\Gamma_2) - \chi(I).
\end{split}
\end{equation*}
By Proposition~\ref{EulerCharacteristic_Proposition}, 
\begin{equation*}
\mathcal{E}(\Gamma_1 \sqcup_I \Gamma_2) = \lambda^{\chi(\Gamma_1 \sqcup_I \Gamma_2)} = \lambda^{\chi(\Gamma_1) + \chi(\Gamma_2) - \chi(I)} = \frac{\lambda^{\chi(\Gamma_1)}\lambda^{\chi(\Gamma_2)}}{\lambda^{\chi(I)}} = \frac{\mathcal{E}(\Gamma_1)\mathcal{E}(\Gamma_2)}{\mathcal{E}(I)},
\end{equation*}
as we expected.
\end{proof}


\subsection{Interface Gluing for the Even Laplacian} \label{EvenLaplacianGluingInterface_Subsection}
In Theorem~\ref{EvenLaplacianGluing_Theorem}, we derive the even Laplacian interface gluing formula. We will consider five different cases by using the indices of the entries of the Laplacian matrix, to identify the positions of the corresponding vertices before and after interface gluing.
\begin{theorem}\label{EvenLaplacianGluing_Theorem}
Let $\Gamma_1$ and $\Gamma_2$ be two graphs, $I$ be the interface, and $\Gamma = \Gamma_1 \sqcup_I \Gamma_2$. Let $V(\Gamma_1) = \{v_1, v_2, \ldots, v_{n-q+1}, \ldots, v_n\}$, $V(\Gamma_2) = \{v_{n-q+1}, \ldots, v_n, \ldots, v_m\}$, and $V(I)=\{v_{n-q+1}, \ldots, v_n\}$. Then, $\Delta^{+}_{\Gamma}$ is given by

\begin{equation*}
\Delta^{+}_{\Gamma}(i,j) = \begin{cases}
\Delta^{+}_{\Gamma_1}(i,j) & \text{if }
\begin{cases}
i \leq n, j < (n-q+1)\\
j \leq n, i < (n-q+1)\\
\end{cases} \text{ $(1)$}\\
\Delta^{+}_{\Gamma_2}(i-n+q,j-n+q) & \text{if }
\begin{cases}
i > n, j \geq (n-q+1)\\
j > n, i \geq (n-q+1)\\
\end{cases} \text{ $(2)$}\\
\Delta^{+}_{\Gamma_1}(i,i)+\Delta^{+}_{\Gamma_2}(i-n+q,i-n+q)-\mu & \text{if } (n-q+1) \leq i=j \leq n \text{ $(3)$}\\
-\Delta^{+}_{\Gamma_1}(i,j) \times \Delta^{+}_{\Gamma_2}(i-n+q,j-n+q) & \text{if } (n-q+1) \leq i \neq j \leq n \text{ $(4)$}\\
0 & \text{otherwise,} \text{ $(5)$}
\end{cases}
\end{equation*}
where $\mu=\sum_{j=n-q+1,j \neq i}^n |\Delta^{+}_{\Gamma_1}(i,j)|$ is the number of neighbors of $v_i$ in the interface.
\end{theorem}

\begin{proof}
We start from verifying (1), and suppose $i \neq j$. If $i \leq n, j < (n-q+1)$ or $j \leq n, i < (n-q+1)$, then both $v_i$ and $v_j$ are in $V(\Gamma_1)$, and at least one of them is not in $V(I)$. Note that interface gluing requires the interface $I$ to be a subgraph of $\Gamma_1$ and $\Gamma_2$, therefore no new edges are created after gluing. Thus, if $v_i$ and $v_j$ were adjacent before gluing, $\Delta^{+}_{\Gamma}(i,j)=\Delta^{+}_{\Gamma_1}(i,j)=-1$, and otherwise $\Delta^{+}_{\Gamma}(i,j)=\Delta^{+}_{\Gamma_1}(i,j)=0$. Now, we suppose $i=j$, then $v_i \in V(\Gamma_1)$ and $v_i \notin V(I)$. It follows that $\text{val}(v_i)$ does not change after gluing. Hence, $\Delta^{+}_{\Gamma}(i,i)=\Delta^{+}_{\Gamma_1}(i,i)=\text{val}(v_i)$.

We verify (2) in a similar way. Suppose $i \neq j$. If $i > n, j \geq (n-q+1)$ or $j > n, i \geq (n-q+1)$, then both $v_i$ and $v_j$ are in $V(\Gamma_2)$, and at least one of them is not in the interface $V(I)$. Then, if $v_i$ and $v_j$ were adjacent before gluing, $\Delta^{+}_{\Gamma}(i,j)=\Delta^{+}_{\Gamma_2}(i-n+q,j-n+q)=-1$, and otherwise $\Delta^{+}_{\Gamma}(i,j)=\Delta^{+}_{\Gamma_2}(i-n+q,j-n+q)=0$. Now, we suppose $i=j$, then $v_i \in V(\Gamma_2)$ and $v_i \notin V(I)$. Therefore, $\Delta^{+}_{\Gamma}(i,i)=\Delta^{+}_{\Gamma_2}(i-n+q,i-n+q)=\text{val}(v_i)$. The reason for having $(i-n+q,j-n+q)$ is that the first $(n-q)$ vertices in $\Gamma_1 \sqcup_I \Gamma_2$ only belong to $\Gamma_1$.

Now, we verify (3). When $(n-q+1) \leq i=j \leq n$, we have $v_i \in V(I)$, and the number of neighbors of $v_i$ in $\Gamma_1 \sqcup_I \Gamma_2$ equals the sum of its neighbors in $\Gamma_1$ and $\Gamma_2$, minus the number of neighbors in $I$, to account for double-counting. Thus, we have $\Delta^{+}_{\Gamma}(i,i)=\Delta^{+}_{\Gamma_1}(i,i)+\Delta^{+}_{\Gamma_2}(i-n+1,i-n+q)-\mu$, where $\mu=\sum_{j=n-q+1,j \neq i}^n |\Delta^{+}_{\Gamma_1}(i,j)|$. The reason for having $(i-n+q)$ is the same as before.

Now, we verify (4). When $(n-q+1) \leq i \neq j \leq n$, we have both $v_i$ and $v_j$ are in $V(I)$. If $v_i$ and $v_j$ are not adjacent before and after gluing, then $\Delta^{+}_{\Gamma}(i,j)=-\Delta^{+}_{\Gamma_1}(i,j) \times \Delta^{+}_{\Gamma_2}(i-n+q,j-n+q)=-0 \times 0=0$; otherwise, $\Delta^{+}_{\Gamma}(i,j)=-\Delta^{+}_{\Gamma_1}(i,j) \times \Delta^{+}_{\Gamma_2}(i-n+q,j-n+q)=-(-1) \times (-1)=-1$.

Finally, we verify (5). If none of (1), (2), (3), or (4) holds, we have one of $v_i$ and $v_j$ is in $V(\Gamma_1)$, the other is in $V(\Gamma_2)$, and neither of them is in $V(I)$. Thus, they cannot be adjacent in $\Gamma_1 \sqcup_I \Gamma_2$, so $\Delta^{+}_{\Gamma}(i,j)=0$.
\end{proof}

\begin{remark}
Observe that the interface gluing formula for the even Laplacian is independent of graph orientation. This is attributed to the fact that the even Laplacian describes how vertices are connected within a graph, and that vertices do not have an orientation.
\end{remark}


\subsection{Interface Gluing for the Odd Laplacian} \label{OddLaplacianGluingInterface_Subsection}

For the interface gluing of odd Laplacians, we will discuss two cases. In Theorem~\ref{OddLaplacianGluing_Theorem}, we derive the odd Laplacian gluing formula when the interface $I$ contains both vertices and edges. We specialize this formula to the case where $I$ only contains vertices in Corollary~\ref{OddInterfaceGluing_Corollary}. Prior to this derivation, it will be helpful to formulate Proposition~\ref{OddLaplacian_Proposition} as a basis for generating the odd Laplacian matrix for connected oriented graphs.

\begin{proposition}\label{OddLaplacian_Proposition}
Let $\Gamma$ be an oriented graph with at least one edge. The odd Laplacian $\Delta^-_{\Gamma}$ is given by
$$\Delta^-_{\Gamma}(i,j) := \begin{cases}
2 & i=j\\
0 & i \neq j \text{, } e_i \text{ is not incident to } e_j\\
1 & i \neq j \text{, } e_i \text{ is incident to } e_j \text{ at } v_s \text{, and } e_i \text{, } e_j \text{ both start or end at } v_s\\
-1 & i \neq j \text{, } e_i \text{ is incident to } e_j \text{ at } v_s \text{, one of the edges starts at } v_s \text{, the other ends at } v_s.
\end{cases}$$
\end{proposition}

\begin{proof}
Let $\mathcal{I}_{\Gamma}$ be the incidence matrix of $\Gamma$, and write $\mathcal{I}_{\Gamma}$ in the form of a combination of column vectors, $\mathcal{I}_{\Gamma}=(c_1,c_2,\ldots,c_n)$, where each $c_i$ is an $n \times 1$ vector with exactly one $-1$, exactly one $1$, and all others are $0$. Since $\Delta^-_{\Gamma}=\mathcal{I}_{\Gamma}^t \mathcal{I}_{\Gamma}$, it follows that $\Delta^-_{\Gamma}(i,j)=c_i^t c_j$. If $i=j$, $\Delta^-_{\Gamma}(i,j)=c_i^t c_i=(-1)^2+1^2=2$.

For any two distinct edges $e_i$ and $e_j$ in $\Gamma$, if $e_i$ is not incident to $e_j$, then we have $\Delta^-_{\Gamma}(i,j)=c_i^t c_j=0$. If $e_i$ is incident to $e_j$ at $v_s$, and $v_s$ is the start point of both edges (or is the end point of both edges), then we have $\Delta^-_{\Gamma}(i,j)=c_i^t c_j=\mathcal{I}_{\Gamma}(s,i)\mathcal{I}_{\Gamma}(s,j)$, where $\mathcal{I}_{\Gamma}(s,i)$ and $\mathcal{I}_{\Gamma}(s,j)$ are both $-1$ or both $1$. Hence we have $\Delta^-_{\Gamma}(i,j)=1$. Similarly, if $e_i$ is incident to $e_j$ at $v_s$, and $v_s$ is the start point of one edge, and is the end point of the other, then we have $\Delta^-_{\Gamma}(i,j)=c_i^t c_j=\mathcal{I}_{\Gamma}(s,i)\mathcal{I}_{\Gamma}(s,j)$, where one of $\mathcal{I}_{\Gamma}(s,i)$ and $\mathcal{I}_{\Gamma}(s,j)$ is $-1$, and the other is $1$. Hence we have $\Delta^-_{\Gamma}(i,j)=-1$.
\end{proof}

\begin{remark}
The odd Laplacian of a graph with only isolated vertices is a $0 \times 0$-matrix whose characteristic polynomial is 1.
\end{remark}

\begin{theorem} \label{OddLaplacianGluing_Theorem}
Let $\Gamma_1$ and $\Gamma_2$ be two oriented graphs with 
\begin{eqnarray*}
V(\Gamma_1) = \{v_1, v_2, \ldots, v_{n-q+1}, \ldots, v_n\}, & V(\Gamma_2) = \{v_{n-q+1}, \ldots, v_n, \ldots, v_m\},\\
E(\Gamma_1) = \{e_1,e_2,\ldots,e_{p-r+1},\ldots,e_p\}, & E(\Gamma_2) = \{e_{p-r+1},\ldots,e_p,\ldots,e_t\}. 
\end{eqnarray*}
Let $I$ be an interface of $\Gamma_1$ and $\Gamma_2$ such that
\begin{eqnarray*}
V(I) = \{v_{n-q+1}, \ldots, v_n\}, &
E(I) = \{e_{p-r+1}, \ldots, e_p\}. 
\end{eqnarray*}
Let $\Gamma = \Gamma_1 \sqcup_I \Gamma_2$, then, $\Delta^{-}_{\Gamma}$ is given by
\begin{equation*}
\Delta^{-}_{\Gamma}(i,j) = \begin{cases}
\Delta^{-}_{\Gamma_1}(i,j) & \text{if } i,j \leq p \\ 
\Delta^{-}_{\Gamma_2}(i-p+r,j-p+r) & \text{if } i,j \geq (p-r+1) \\
\zeta & \text{otherwise,}
\end{cases}
\end{equation*}
where 
\begin{equation*}
\zeta=\begin{cases}
0 & \text{if } e_i \text{ is not incident to } e_j\\ 
1 & \text{if } e_i \text{ is incident to } e_j \text{ at } v_s \text{, and } e_i \text{ and } e_j \text{ both start or end at } v_s\\
-1 & \text{if } e_i \text{ is incident to } e_j \text{ at } v_s \text{, and one of the edges starts at } v_s \text{, the other ends at } v_s.
\end{cases}
\end{equation*}
\end{theorem}

\begin{proof}
Note that interface gluing does not create new edges. Thus $E(\Gamma)=E(\Gamma_1) \cup E(\Gamma_2)=\{e_1,e_2,\ldots,e_t\}$, and $|E(\Gamma)|=|E(\Gamma_1)|+|E(\Gamma_2)|-|E(I)|$. By Proposition~\ref{OddLaplacian_Proposition}, the entries in an odd Laplacian are determined by the orientations of the corresponding edges.

When $i \leq p$ and $j \leq p$, it follows that $e_i$ and $e_j$ are both in $E(\Gamma_1)$. Their $\zeta$-condition and orientations are not changed after gluing, thus $\Delta^{-}_{\Gamma}(i,j)=\Delta^{-}_{\Gamma_1}(i,j)$. Similarly, when $i \geq (p-r+1)$ and $j \geq (p-r+1)$, it holds that $e_i$ and $e_j$ are both in $E(\Gamma_2)$. Their $\zeta$-condition and orientations are not changed after gluing, thus $\Delta^{-}_{\Gamma}(i,j)=\Delta^{-}_{\Gamma_2}(i-p+r,j-p+r)$. The reason for having $(i-p+r,j-p+r)$ is that the first $(p-r)$ edges in $\Gamma_1 \sqcup_I \Gamma_2$ only belong to $\Gamma_1$. When $i \geq (p-r+1), j \leq p$, we have that $e_i$ is in $E(\Gamma_2)$, and $e_j$ is in $E(\Gamma_1)$. Their orientations are not changed after gluing, however a priori their $\zeta$-condition could change. By Proposition~\ref{OddLaplacian_Proposition}, we obtain the corresponding value of $\zeta$ for all three $\zeta$-conditions between $e_i$ and $e_j$. The resulting value of $\zeta $ also works when $i \leq p, j \geq (p-r+1)$, and this concludes the proof.
\end{proof}

In the following corollary we consider the case in which the interface subgraph has no edges.
\begin{corollary}\label{OddInterfaceGluing_Corollary}
Let $\Gamma_1$ and $\Gamma_2$ be two oriented graphs, $\{e_1,e_2,\ldots,e_p\}$ be the edges of $\Gamma_1$, and $\{e_{p+1},e_{p+2},\ldots,e_t\}$ be the edges of $\Gamma_2$. Let the interface $I$ be a set of $q$ vertices $\{v_{n-q+1}, \ldots, v_n\}$. Let $\Gamma = \Gamma_1 \sqcup_I \Gamma_2$, then, $\Delta^{-}_{\Gamma}$ is a block matrix given by
\begin{equation*}
\Delta^{-}_{\Gamma_1 \sqcup_I \Gamma_2} = \begin{bmatrix}
\Delta^{-}_{\Gamma_1} & Q^t \\
Q & \Delta^{-}_{\Gamma_2}
\end{bmatrix},
\end{equation*}
where 
\begin{equation*} Q(i,j) = \begin{cases}
0 & \text{if } e_{p+i} \text{ is not incident to } e_j\\ 
1 & \text{if } e_{p+i} \text{ is incident to } e_j \text{ at } v_s \text{, and } e_{p+i} \text{ and } e_j \text{ both start or end at } v_s\\
-1 & \text{if } e_{p+i} \text{ is incident to } e_j \text{ at } v_s \text{, and one of the edges starts at } v_s \text{, the other ends at } v_s.
\end{cases}
\end{equation*}
\end{corollary}

\begin{proof}
Since $I$ only contains vertices, we have a particular case of Theorem~\ref{OddLaplacianGluing_Theorem} where $r=0$. Therefore, we automatically have $\Delta^{-}_{\Gamma}(i,j)=\Delta^{-}_{\Gamma_1}(i,j)$ when $i \leq p$ and $j \leq p$, and $\Delta^{-}_{\Gamma}(i,j)=\Delta^{-}_{\Gamma_2}(i-p,j-p)$ when $i \geq (p+1)$ and $j \geq (p+1)$. Moreover, when $i \geq (p+1)$ and $j \leq p$, we have that $e_i \in E(\Gamma_2)$ and $e_j \in E(\Gamma_1)$. The $\zeta$ terms in Theorem~\ref{OddLaplacianGluing_Theorem} now form the block $Q$ for all three $\zeta$-conditions between $e_i$ and $e_j$ in Proposition~\ref{OddLaplacian_Proposition}. Similarly, we get the block $Q^t$ when $i \leq p$ and $j \geq (p+1)$.
\end{proof}


\subsection{Interface Gluing Spectra} \label{InterfaceGluingSpectra_Subsection}

We now discuss two explicit interface gluing formulae for the spectrum, one where the interface is a single vertex between two unoriented graphs and another where we add an edge to a connected unoriented graph. The second case can be interpreted as an interface gluing, where we glue the original graph and $P_2$ (representing the added edge) over the interface $I$ where $V(I) = V(P_2)$.


\begin{theorem} \label{VertexInterfaceSpectrum_Theorem}
Let $\Gamma_1$ and $\Gamma_2$ be two arbitrary graphs with $m$ and $n$ vertices respectively, and choose one vertex from each graph to define an interface $I$. Order the vertices in each graph such that the chosen vertex in $\Gamma_1$ is $v^1_m$ and that in $\Gamma_2$ is $v^2_1$. Then, the characteristic polynomial of $\Delta_{\Gamma_1 \sqcup_I \Gamma_2}$, where $V(I) = \{v^1_m\} = \{v^2_1\}$ and $E(I) = \emptyset$, is
\begin{equation*}
p_{\Delta_{\Gamma_1 \sqcup_I \Gamma_2}}(\lambda) = p_{\Delta_{\Gamma_1}}(\lambda) p_{\Delta_{\Gamma_2 \left(v^2_1, v^2_1\right)}}(\lambda) + p_{\Delta_{\Gamma_1 \left(v^1_m, v^1_m\right)}}(\lambda) p_{\Delta_{\Gamma_2}}(\lambda) + \lambda p_{\Delta_{\Gamma_1 \left(v^1_m, v^1_m\right)}}(\lambda) p_{\Delta_{\Gamma_2 \left(v^2_1, v^2_1\right)}}(\lambda).
\end{equation*}
\end{theorem}

\begin{proof}
The graph Laplacian matrix $\Delta_{\Gamma_1 \sqcup_I \Gamma_2}$ takes the following form: \footnote{We denote by $M^{(i,j)}$ the $(i,j)$-index of matrix $M$}

\begin{equation*}
\Delta_{\Gamma_1 \sqcup_I \Gamma_2} = \begin{bmatrix}
\Delta^{(1,1)}_{\Gamma_1} & \hdots & \Delta^{(1,m-1)}_{\Gamma_1} & \Delta^{(1,m)}_{\Gamma_1} & 0 & \hdots & 0 \\
\vdots & \ddots & \vdots & \vdots & 0 & \hdots & 0 \\
\Delta^{(m-1,1)}_{\Gamma_1} & \hdots & \Delta^{(m-1,m-1)}_{\Gamma_1} & \Delta^{(m-1,m)}_{\Gamma_1} & 0 & \hdots & 0 \\
\Delta^{(m,1)}_{\Gamma_1} & \hdots & \Delta^{(m,m-1)}_{\Gamma_1} & \Delta^{(m,m)}_{\Gamma_1} + \Delta^{(1,1)}_{\Gamma_2} & \Delta^{(1,2)}_{\Gamma_2} & \hdots & \Delta^{(1,n)}_{\Gamma_2} \\
0 & \hdots & 0 & \Delta^{(2,1)}_{\Gamma_2} & \Delta^{(2,2)}_{\Gamma_2} & \hdots & \Delta^{(2,n)}_{\Gamma_2} \\
0 & \hdots & 0 & \vdots & \vdots & \ddots & \vdots \\
0 & \hdots & 0 & \Delta^{(n,1)}_{\Gamma_2} & \Delta^{(n,2)}_{\Gamma_2} & \hdots & \Delta^{(n,n)}_{\Gamma_2} \\
\end{bmatrix}.
\end{equation*}

The characteristic polynomial can be factored in two terms by Lemma~\ref{SplitColumn_Lemma}.

\begin{equation*}
\begin{split}
p_{\Delta_{\Gamma_1 \sqcup_I \Gamma_2}}(\lambda) & = \det\begin{bmatrix}
\Delta^{(1,1)}_{\Gamma_1} - \lambda & \hdots & \Delta^{(1,m-1)}_{\Gamma_1} & \Delta^{(1,m)}_{\Gamma_1} & 0 & \hdots & 0 \\
\vdots & \ddots & \vdots & \vdots & 0 & \hdots & 0 \\
\Delta^{(m-1,1)}_{\Gamma_1} & \hdots & \Delta^{(m-1,m-1)}_{\Gamma_1} - \lambda & \Delta^{(m-1,m)}_{\Gamma_1} & 0 & \hdots & 0 \\
\Delta^{(m,1)}_{\Gamma_1} & \hdots & \Delta^{(m,m-1)}_{\Gamma_1} & \Delta^{(m,m)}_{\Gamma_1} + \Delta^{(1,1)}_{\Gamma_2} - \lambda & \Delta^{(1,2)}_{\Gamma_2} & \hdots & \Delta^{(1,n)}_{\Gamma_2} \\
0 & \hdots & 0 & \Delta^{(2,1)}_{\Gamma_2} & \Delta^{(2,2)}_{\Gamma_2} - \lambda & \hdots & \Delta^{(2,n)}_{\Gamma_2} \\
0 & \hdots & 0 & \vdots & \vdots & \ddots & \vdots \\
0 & \hdots & 0 & \Delta^{(n,1)}_{\Gamma_2} & \Delta^{(n,2)}_{\Gamma_2} & \hdots & \Delta^{(n,n)}_{\Gamma_2} - \lambda \\
\end{bmatrix} \\
& = \det\begin{bmatrix}
\Delta^{(1,1)}_{\Gamma_1} - \lambda & \hdots & \Delta^{(1,m-1)}_{\Gamma_1} & \Delta^{(1,m)}_{\Gamma_1} & 0 & \hdots & 0 \\
\vdots & \ddots & \vdots & \vdots & 0 & \hdots & 0 \\
\Delta^{(m-1,1)}_{\Gamma_1} & \hdots & \Delta^{(m-1,m-1)}_{\Gamma_1} - \lambda & \Delta^{(m-1,m)}_{\Gamma_1} & 0 & \hdots & 0 \\
\Delta^{(m,1)}_{\Gamma_1} & \hdots & \Delta^{(m,m-1)}_{\Gamma_1} & \Delta^{(m,m)}_{\Gamma_1} - \lambda & \Delta^{(1,2)}_{\Gamma_2} & \hdots & \Delta^{(1,n)}_{\Gamma_2} \\
0 & \hdots & 0 & 0 & \Delta^{(2,2)}_{\Gamma_2} - \lambda & \hdots & \Delta^{(2,n)}_{\Gamma_2} \\
0 & \hdots & 0 & \vdots & \vdots & \ddots & \vdots \\
0 & \hdots & 0 & 0 & \Delta^{(n,2)}_{\Gamma_2} & \hdots & \Delta^{(n,n)}_{\Gamma_2} - \lambda \\
\end{bmatrix} \\
& + \det\begin{bmatrix}
\Delta^{(1,1)}_{\Gamma_1} - \lambda & \hdots & \Delta^{(1,m-1)}_{\Gamma_1} & 0 & 0 & \hdots & 0 \\
\vdots & \ddots & \vdots & \vdots & 0 & \hdots & 0 \\
\Delta^{(m-1,1)}_{\Gamma_1} & \hdots & \Delta^{(m-1,m-1)}_{\Gamma_1} - \lambda & 0 & 0 & \hdots & 0 \\
\Delta^{(m,1)}_{\Gamma_1} & \hdots & \Delta^{(m,m-1)}_{\Gamma_1} & \Delta^{(1,1)}_{\Gamma_2} & \Delta^{(1,2)}_{\Gamma_2} & \hdots & \Delta^{(1,n)}_{\Gamma_2} \\
0 & \hdots & 0 & \Delta^{(2,1)}_{\Gamma_2} & \Delta^{(2,2)}_{\Gamma_2} - \lambda & \hdots & \Delta^{(2,n)}_{\Gamma_2} \\
0 & \hdots & 0 & \vdots & \vdots & \ddots & \vdots \\
0 & \hdots & 0 & \Delta^{(n,1)}_{\Gamma_2} & \Delta^{(n,2)}_{\Gamma_2} & \hdots & \Delta^{(n,n)}_{\Gamma_2} - \lambda \\
\end{bmatrix}. \\
\end{split}
\end{equation*}

The first determinant is a block upper triangular matrix whereas the second is block lower triangular, so $p_{\Delta_{\Gamma_1 \sqcup_I \Gamma_2}}(\lambda)$ can be simplified by Lemma~\ref{BlockTriangular_Lemma},

\begin{equation*}
\begin{split}
p_{\Delta_{\Gamma_1 \sqcup_I \Gamma_2}}(\lambda)
& = \det\begin{bmatrix}
\Delta^{(1,1)}_{\Gamma_1} - \lambda & \hdots & \Delta^{(1,m)}_{\Gamma_1} \\
\vdots & \ddots & \vdots \\
\Delta^{(m,1)}_{\Gamma_1} & \hdots & \Delta^{(m,m)}_{\Gamma_1} - \lambda
\end{bmatrix} \det\begin{bmatrix}
\Delta^{(2,2)}_{\Gamma_2} - \lambda & \hdots & \Delta^{(2,n)}_{\Gamma_2} \\
\vdots & \ddots & \vdots \\
\Delta^{(n,2)}_{\Gamma_2} & \hdots & \Delta^{(n,n)}_{\Gamma_2} - \lambda
\end{bmatrix} \\
& + \det\begin{bmatrix}
\Delta^{(1,1)}_{\Gamma_1} - \lambda & \hdots & \Delta^{(1,m-1)}_{\Gamma_1} \\
\vdots & \ddots & \vdots \\
\Delta^{(m-1,1)}_{\Gamma_1} & \hdots & \Delta^{(m-1,m-1)}_{\Gamma_1} - \lambda \\
\end{bmatrix} \det\begin{bmatrix}
\Delta^{(1,1)}_{\Gamma_2} & \Delta^{(1,2)}_{\Gamma_2} & \hdots & \Delta^{(1,n)}_{\Gamma_2} \\
\Delta^{(2,1)}_{\Gamma_2} & \Delta^{(2,2)}_{\Gamma_2} - \lambda & \hdots & \Delta^{(2,n)}_{\Gamma_2} \\
\vdots & \vdots & \ddots & \vdots \\
\Delta^{(n,1)}_{\Gamma_2} & \Delta^{(n,2)}_{\Gamma_2} & \hdots & \Delta^{(n,n)}_{\Gamma_2} - \lambda \\
\end{bmatrix} \\
& = p_{\Delta_{\Gamma_1}}(\lambda) p_{\Delta_{\Gamma_2 \left(v^2_1, v^2_1\right)}}(\lambda) + p_{\Delta_{\Gamma_1 \left(v^1_m, v^1_m\right)}}(\lambda) \det\begin{bmatrix}
\Delta^{(1,1)}_{\Gamma_2} & \Delta^{(1,2)}_{\Gamma_2} & \hdots & \Delta^{(1,n)}_{\Gamma_2} \\
\Delta^{(2,1)}_{\Gamma_2} & \Delta^{(2,2)}_{\Gamma_2} - \lambda & \hdots & \Delta^{(2,n)}_{\Gamma_2} \\
\vdots & \vdots & \ddots & \vdots \\
\Delta^{(n,1)}_{\Gamma_2} & \Delta^{(n,2)}_{\Gamma_2} & \hdots & \Delta^{(n,n)}_{\Gamma_2} - \lambda \\
\end{bmatrix},
\end{split}
\end{equation*}

where the subscripts $\left(v^1_m, v^1_m\right)$ and $\left(v^2_1, v^2_1\right)$ emerge because the minors of the Laplacian matrices depend on which two vertices are glued. The remaining determinant can be solved explicitly by using Lemma~\ref{SplitColumn_Lemma} and cofactor expansion,

\begin{equation*}
\begin{split}
\det\begin{bmatrix}
\Delta^{(1,1)}_{\Gamma_2} & \Delta^{(1,2)}_{\Gamma_2} & \hdots & \Delta^{(1,n)}_{\Gamma_2} \\
\Delta^{(2,1)}_{\Gamma_2} & \Delta^{(2,2)}_{\Gamma_2} - \lambda & \hdots & \Delta^{(2,n)}_{\Gamma_2} \\
\vdots & \vdots & \ddots & \vdots \\
\Delta^{(n,1)}_{\Gamma_2} & \Delta^{(n,2)}_{\Gamma_2} & \hdots & \Delta^{(n,n)}_{\Gamma_2} - \lambda \\
\end{bmatrix}
& = \det\begin{bmatrix}
\Delta^{(1,1)}_{\Gamma_2} - \lambda & \Delta^{(1,2)}_{\Gamma_2} & \hdots & \Delta^{(1,n)}_{\Gamma_2} \\
\Delta^{(2,1)}_{\Gamma_2} & \Delta^{(2,2)}_{\Gamma_2} - \lambda & \hdots & \Delta^{(2,n)}_{\Gamma_2} \\
\vdots & \vdots & \ddots & \vdots \\
\Delta^{(n,1)}_{\Gamma_2} & \Delta^{(n,2)}_{\Gamma_2} & \hdots & \Delta^{(n,n)}_{\Gamma_2} - \lambda \\
\end{bmatrix} \\
& + \det\begin{bmatrix}
\lambda & \Delta^{(1,2)}_{\Gamma_2} & \hdots & \Delta^{(1,n)}_{\Gamma_2} \\
0 & \Delta^{(2,2)}_{\Gamma_2} - \lambda & \hdots & \Delta^{(2,n)}_{\Gamma_2} \\
\vdots & \vdots & \ddots & \vdots \\
0 & \Delta^{(n,2)}_{\Gamma_2} & \hdots & \Delta^{(n,n)}_{\Gamma_2} - \lambda \\
\end{bmatrix} \\
& = p_{\Delta_{\Gamma_2}}(\lambda) + \lambda \det\begin{bmatrix}
\Delta^{(2,2)}_{\Gamma_2} - \lambda & \hdots & \Delta^{(2,n)}_{\Gamma_2} \\
\vdots & \ddots & \vdots \\
\Delta^{(n,2)}_{\Gamma_2} & \hdots & \Delta^{(n,n)}_{\Gamma_2} - \lambda \\
\end{bmatrix} \\
& = p_{\Delta_{\Gamma_2}}(\lambda) + \lambda p_{\Delta_{\Gamma_2 \left(v^2_1, v^2_1\right)}}(\lambda).
\end{split}
\end{equation*}

Therefore, $p_{\Delta_{\Gamma_1 \sqcup_I \Gamma_2}}$ is given to be

\begin{equation*}
\begin{split}
p_{\Delta_{\Gamma_1 \sqcup_I \Gamma_2}}(\lambda) & = p_{\Delta_{\Gamma_1}}(\lambda) p_{\Delta_{\Gamma_2 \left(v^2_1, v^2_1\right)}}(\lambda) + p_{\Delta_{\Gamma_1 \left(v^1_m, v^1_m\right)}}(\lambda) \left(p_{\Delta_{\Gamma_2}}(\lambda) + \lambda p_{\Delta_{\Gamma_2 \left(v^2_1, v^2_1\right)}}(\lambda)\right) \\
& = p_{\Delta_{\Gamma_1}}(\lambda) p_{\Delta_{\Gamma_2 \left(v^2_1, v^2_1\right)}}(\lambda) + p_{\Delta_{\Gamma_1 \left(v^1_m, v^1_m\right)}}(\lambda) p_{\Delta_{\Gamma_2}}(\lambda) + \lambda p_{\Delta_{\Gamma_1 \left(v^1_m, v^1_m\right)}}(\lambda) p_{\Delta_{\Gamma_2 \left(v^2_1, v^2_1\right)}}(\lambda),
\end{split}
\end{equation*}
as we wanted.
\end{proof}

\begin{remark}
Note that the above result holds only in the case when the interface is composed of one vertex from each graph. An analogous formula for the interface gluing of multiple vertices ($|V(I)| \geq 2$) can be understood by gluing two vertices first before ``internally gluing'' vertices within the same graph. However, the derivation of this formula depends on the degrees of the vertices, and a compact form has yet to be developed.
\end{remark}


We now show how the spectrum changes when a new edge is added between two disconnected vertices of a graph. As mentioned before, this can be interpreted as interface gluing by considering the original graph as one graph and $P_2$ as the other graph.

\begin{theorem} \label{EdgeInterfaceSpectrum_Theorem}
Let $\Gamma_1$ be an unoriented graph with $m$ vertices, and choose two nonadjacent vertices $v_1, v_2 \in V(\Gamma_1)$. Let $I$ be an interface such that $V(I) = \{v_1, v_2\}$, and let $P_2$ be the path graph where $V(P_2) = V(I)$. Then, the characteristic polynomial of $\Delta_{\Gamma_1 \sqcup_I P_2}$ is the following:

\begin{equation*}
\begin{split}
p_{\Delta_{\Gamma_1 \sqcup_I P_2}}(\lambda) & = p_{\Delta_{\Gamma_1}}(\lambda) + p_{\Delta_{\Gamma_1 (v_1, v_1)}}(\lambda) + p_{\Delta_{\Gamma_1 (v_2, v_2)}}(\lambda) - 2 (-1)^{v_1+v_2} \det\bigg((\Delta_{\Gamma_1} - \lambda I)_{(v_1, v_2)}\bigg).
\end{split}
\end{equation*}
\end{theorem}

\begin{proof}
The graph Laplacian of $\Gamma_1 \sqcup_I P_2$ has the following form.

\begin{equation*}
\Delta_{\Gamma_1 \sqcup_I P_2} = \begin{bmatrix}
\Delta_{\Gamma_1}^{(1,1)} & \hdots & \Delta_{\Gamma_1}^{(1,v_1)} & \hdots & \Delta_{\Gamma_1}^{(1,v_2)} & \hdots & \Delta_{\Gamma_1}^{(1,m)} \\
\vdots & \ddots & \vdots & \ddots & \vdots & \ddots & \vdots \\
\Delta_{\Gamma_1}^{(v_1,1)} & \hdots & \Delta_{\Gamma_1}^{(v_1,v_1)} + 1 & \hdots & -1 & \hdots & \Delta_{\Gamma_1}^{(v_1,m)} \\
\vdots & \ddots & \vdots & \ddots & \vdots & \ddots & \vdots \\
\Delta_{\Gamma_1}^{(v_2,1)} & \hdots & -1 & \hdots & \Delta_{\Gamma_1}^{(v_2,v_2)} + 1 & \hdots & \Delta_{\Gamma_1}^{(v_2,m)} \\
\vdots & \ddots & \vdots & \ddots & \vdots & \ddots & \vdots \\
\Delta_{\Gamma_1}^{(m,1)} & \hdots & \Delta_{\Gamma_1}^{(m,v_1)} & \hdots & \Delta_{\Gamma_1}^{(m,v_2)} & \hdots & \Delta_{\Gamma_1}^{(m,m)}
\end{bmatrix}.
\end{equation*}

Its characteristic polynomial is the following:

\begin{equation*}
\begin{split}
p_{\Delta_{\Gamma_1 \sqcup_I P_2}}(\lambda) & = \det \begin{bmatrix}
\Delta_{\Gamma_1}^{(1,1)} - \lambda & \hdots & \Delta_{\Gamma_1}^{(1,v_1)} & \hdots & \Delta_{\Gamma_1}^{(1,v_2)} & \hdots & \Delta_{\Gamma_1}^{(1,m)} \\
\vdots & \ddots & \vdots & \ddots & \vdots & \ddots & \vdots \\
\Delta_{\Gamma_1}^{(v_1,1)} & \hdots & \Delta_{\Gamma_1}^{(v_1,v_1)} + 1 - \lambda & \hdots & -1 & \hdots & \Delta_{\Gamma_1}^{(v_1,m)} \\
\vdots & \ddots & \vdots & \ddots & \vdots & \ddots & \vdots \\
\Delta_{\Gamma_1}^{(v_2,1)} & \hdots & -1 & \hdots & \Delta_{\Gamma_1}^{(v_2,v_2)} + 1 - \lambda & \hdots & \Delta_{\Gamma_1}^{(v_2,m)} \\
\vdots & \ddots & \vdots & \ddots & \vdots & \ddots & \vdots \\
\Delta_{\Gamma_1}^{(m,1)} & \hdots & \Delta_{\Gamma_1}^{(m,v_1)} & \hdots & \Delta_{\Gamma_1}^{(m,v_2)} & \hdots & \Delta_{\Gamma_1}^{(m,m)} - \lambda
\end{bmatrix}. \\
\end{split}
\end{equation*}

We first use Lemma \ref{SplitColumn_Lemma} on the $(v_1)^{th}$ column.

\begin{equation*}
\begin{split}
p_{\Delta_{\Gamma_1 \sqcup_I P_2}}(\lambda) & = \det \begin{bmatrix}
\Delta_{\Gamma_1}^{(1,1)} - \lambda & \hdots & \Delta_{\Gamma_1}^{(1,v_1)} & \hdots & \Delta_{\Gamma_1}^{(1,v_2)} & \hdots & \Delta_{\Gamma_1}^{(1,m)} \\
\vdots & \ddots & \vdots & \ddots & \vdots & \ddots & \vdots \\
\Delta_{\Gamma_1}^{(v_1,1)} & \hdots & \Delta_{\Gamma_1}^{(v_1,v_1)} - \lambda & \hdots & -1 & \hdots & \Delta_{\Gamma_1}^{(v_1,m)} \\
\vdots & \ddots & \vdots & \ddots & \vdots & \ddots & \vdots \\
\Delta_{\Gamma_1}^{(v_2,1)} & \hdots & 0 & \hdots & \Delta_{\Gamma_1}^{(v_2,v_2)} + 1 - \lambda & \hdots & \Delta_{\Gamma_1}^{(v_2,m)} \\
\vdots & \ddots & \vdots & \ddots & \vdots & \ddots & \vdots \\
\Delta_{\Gamma_1}^{(m,1)} & \hdots & \Delta_{\Gamma_1}^{(m,v_1)} & \hdots & \Delta_{\Gamma_1}^{(m,v_2)} & \hdots & \Delta_{\Gamma_1}^{(m,m)} - \lambda
\end{bmatrix} \\
& + \det \begin{bmatrix}
\Delta_{\Gamma_1}^{(1,1)} - \lambda & \hdots & 0 & \hdots & \Delta_{\Gamma_1}^{(1,v_2)} & \hdots & \Delta_{\Gamma_1}^{(1,m)} \\
\vdots & \ddots & \vdots & \ddots & \vdots & \ddots & \vdots \\
\Delta_{\Gamma_1}^{(v_1,1)} & \hdots & 1 & \hdots & -1 & \hdots & \Delta_{\Gamma_1}^{(v_1,m)} \\
\vdots & \ddots & \vdots & \ddots & \vdots & \ddots & \vdots \\
\Delta_{\Gamma_1}^{(v_2,1)} & \hdots & -1 & \hdots & \Delta_{\Gamma_1}^{(v_2,v_2)} + 1 - \lambda & \hdots & \Delta_{\Gamma_1}^{(v_2,m)} \\
\vdots & \ddots & \vdots & \ddots & \vdots & \ddots & \vdots \\
\Delta_{\Gamma_1}^{(m,1)} & \hdots & 0 & \hdots & \Delta_{\Gamma_1}^{(m,v_2)} & \hdots & \Delta_{\Gamma_1}^{(m,m)} - \lambda
\end{bmatrix} \\
& =: X_1(\lambda) + X_2(\lambda).
\end{split}
\end{equation*}

We focus on the first summand $X_1(\lambda)$ initially. We can further split this using Lemma \ref{SplitColumn_Lemma} on the $(v_2)^{th}$ column in the following way:

\begin{equation*}
\begin{split}
X_1(\lambda) & = \det \begin{bmatrix}
\Delta_{\Gamma_1}^{(1,1)} - \lambda & \hdots & \Delta_{\Gamma_1}^{(1,v_1)} & \hdots & \Delta_{\Gamma_1}^{(1,v_2)} & \hdots & \Delta_{\Gamma_1}^{(1,m)} \\
\vdots & \ddots & \vdots & \ddots & \vdots & \ddots & \vdots \\
\Delta_{\Gamma_1}^{(v_1,1)} & \hdots & \Delta_{\Gamma_1}^{(v_1,v_1)} - \lambda & \hdots & 0 & \hdots & \Delta_{\Gamma_1}^{(v_1,m)} \\
\vdots & \ddots & \vdots & \ddots & \vdots & \ddots & \vdots \\
\Delta_{\Gamma_1}^{(v_2,1)} & \hdots & 0 & \hdots & \Delta_{\Gamma_1}^{(v_2,v_2)} - \lambda & \hdots & \Delta_{\Gamma_1}^{(v_2,m)} \\
\vdots & \ddots & \vdots & \ddots & \vdots & \ddots & \vdots \\
\Delta_{\Gamma_1}^{(m,1)} & \hdots & \Delta_{\Gamma_1}^{(m,v_1)} & \hdots & \Delta_{\Gamma_1}^{(m,v_2)} & \hdots & \Delta_{\Gamma_1}^{(m,m)} - \lambda
\end{bmatrix} \\
& + \det \begin{bmatrix}
\Delta_{\Gamma_1}^{(1,1)} - \lambda & \hdots & \Delta_{\Gamma_1}^{(1,v_1)} & \hdots & 0 & \hdots & \Delta_{\Gamma_1}^{(1,m)} \\
\vdots & \ddots & \vdots & \ddots & \vdots & \ddots & \vdots \\
\Delta_{\Gamma_1}^{(v_1,1)} & \hdots & \Delta_{\Gamma_1}^{(v_1,v_1)} - \lambda & \hdots & -1 & \hdots & \Delta_{\Gamma_1}^{(v_1,m)} \\
\vdots & \ddots & \vdots & \ddots & \vdots & \ddots & \vdots \\
\Delta_{\Gamma_1}^{(v_2,1)} & \hdots & 0 & \hdots & 1 & \hdots & \Delta_{\Gamma_1}^{(v_2,m)} \\
\vdots & \ddots & \vdots & \ddots & \vdots & \ddots & \vdots \\
\Delta_{\Gamma_1}^{(m,1)} & \hdots & \Delta_{\Gamma_1}^{(m,v_1)} & \hdots & 0 & \hdots & \Delta_{\Gamma_1}^{(m,m)} - \lambda
\end{bmatrix}. \\
\end{split}
\end{equation*}

We realize that the first term of $X_1(\lambda)$ is the characteristic polynomial of $\Delta_{\Gamma_1}$. The second term may be simplified using cofactor expansion along the $(v_2)^{th}$ column. Thus,

\begin{equation*}
\begin{split}
X_1(\lambda) & = p_{\Delta_{\Gamma_1}}(\lambda) + \det \begin{bmatrix}
\Delta_{\Gamma_1}^{(1,1)} - \lambda & \hdots & \Delta_{\Gamma_1}^{(1,v_1)} & \hdots & \Delta_{\Gamma_1}^{(1,m)} \\
\vdots & \ddots & \vdots & \ddots & \vdots \\
\Delta_{\Gamma_1}^{(v_1,1)} & \hdots & \Delta_{\Gamma_1}^{(v_1,v_1)} - \lambda & \hdots & \Delta_{\Gamma_1}^{(v_1,m)} \\
\vdots & \ddots & \vdots & \ddots & \vdots \\
\Delta_{\Gamma_1}^{(m,1)} & \hdots & \Delta_{\Gamma_1}^{(m,v_1)} & \hdots & \Delta_{\Gamma_1}^{(m,m)} - \lambda
\end{bmatrix} \\
& + (-1) (-1)^{v_1 + v_2} \det \begin{bmatrix}
\Delta_{\Gamma_1}^{(1,1)} - \lambda & \hdots & \Delta_{\Gamma_1}^{(1,v_1)} & \hdots & \Delta_{\Gamma_1}^{(1,m)} \\
\vdots & \ddots & \vdots & \ddots & \vdots \\
\Delta_{\Gamma_1}^{(v_2,1)} & \hdots & 0 & \hdots & \Delta_{\Gamma_1}^{(v_2,m)} \\
\vdots & \ddots & \vdots & \ddots & \vdots \\
\Delta_{\Gamma_1}^{(m,1)} & \hdots & \Delta_{\Gamma_1}^{(m,v_1)} & \hdots & \Delta_{\Gamma_1}^{(m,m)} - \lambda
\end{bmatrix}. \\
\end{split}
\end{equation*}

We realize that the determinants may be written in terms of the minors of $\Delta_{\Gamma_1} - \lambda I$; namely,

\begin{equation*}
\begin{split}
X_1(\lambda) & = p_{\Delta_{\Gamma_1}}(\lambda) + \det\left(
\left(\Delta_{\Gamma_1} - \lambda I\right)_{(v_2, v_2)}
\right) + (-1)^{v_1 + v_2} (-1) \det\left(
\left(\Delta_{\Gamma_1} - \lambda I\right)_{(v_1, v_2)}
\right) \\
& = p_{\Delta_{\Gamma_1}}(\lambda) + p_{\Delta_{\Gamma_1 (v_2, v_2)}}(\lambda) - (-1)^{v_1 + v_2} \det\left(
\left(\Delta_{\Gamma_1} - \lambda I\right)_{(v_1, v_2)}
\right).
\end{split}
\end{equation*}

We now turn our attention to the second term $X_2(\lambda)$. It follows from cofactor expansion on the $(v_1)^{th}$ column that

\begin{equation*}
\begin{split}
X_2(\lambda) & = \det \begin{bmatrix}
\Delta_{\Gamma_1}^{(1,1)} - \lambda & \hdots & \Delta_{\Gamma_1}^{(1,v_2)} & \hdots & \Delta_{\Gamma_1}^{(1,m)} \\
\vdots & \ddots & \vdots & \ddots & \vdots \\
\Delta_{\Gamma_1}^{(v_2,1)} & \hdots & \Delta_{\Gamma_1}^{(v_2,v_2)} + 1 - \lambda & \hdots & \Delta_{\Gamma_1}^{(v_2,m)} \\
\vdots & \ddots & \vdots & \ddots & \vdots \\
\Delta_{\Gamma_1}^{(m,1)} & \hdots & \Delta_{\Gamma_1}^{(m,v_2)} & \hdots & \Delta_{\Gamma_1}^{(m,m)} - \lambda
\end{bmatrix} \\
& + (-1) (-1)^{v_2 + v_1} \det \begin{bmatrix}
\Delta_{\Gamma_1}^{(1,1)} - \lambda & \hdots & \Delta_{\Gamma_1}^{(1,v_2)} & \hdots & \Delta_{\Gamma_1}^{(1,m)} \\
\vdots & \ddots & \vdots & \ddots & \vdots \\
\Delta_{\Gamma_1}^{(v_1,1)} & \hdots & -1 & \hdots & \Delta_{\Gamma_1}^{(v_1,m)} \\
\vdots & \ddots & \vdots & \ddots & \vdots \\
\Delta_{\Gamma_1}^{(m,1)} & \hdots & \Delta_{\Gamma_1}^{(m,v_2)} & \hdots & \Delta_{\Gamma_1}^{(m,m)} - \lambda
\end{bmatrix} \\
& =: X_{21}(\lambda) - (-1)^{v_2 + v_1} X_{22}(\lambda).
\end{split}
\end{equation*}

We use Lemma \ref{SplitColumn_Lemma} on the $(v_2)^{th}$ column of $X_{21}(\lambda)$ to obtain the following.

\begin{equation*}
\begin{split}
X_{21}(\lambda) & = \det \begin{bmatrix}
\Delta_{\Gamma_1}^{(1,1)} & \hdots & \Delta_{\Gamma_1}^{(1,v_2)} & \hdots & \Delta_{\Gamma_1}^{(1,m)} \\
\vdots & \ddots & \vdots & \ddots & \vdots \\
\Delta_{\Gamma_1}^{(v_2,1)} & \hdots & \Delta_{\Gamma_1}^{(v_2,v_2)} & \hdots & \Delta_{\Gamma_1}^{(v_2,m)} \\
\vdots & \ddots & \vdots & \ddots & \vdots \\
\Delta_{\Gamma_1}^{(m,1)} & \hdots & \Delta_{\Gamma_1}^{(m,v_2)} & \hdots & \Delta_{\Gamma_1}^{(m,m)}
\end{bmatrix} + \det \begin{bmatrix}
\Delta_{\Gamma_1}^{(1,1)} & \hdots & 0 & \hdots & \Delta_{\Gamma_1}^{(1,m)} \\
\vdots & \ddots & \vdots & \ddots & \vdots \\
\Delta_{\Gamma_1}^{(v_2,1)} & \hdots & 1 & \hdots & \Delta_{\Gamma_1}^{(v_2,m)} \\
\vdots & \ddots & \vdots & \ddots & \vdots \\
\Delta_{\Gamma_1}^{(m,1)} & \hdots & 0 & \hdots & \Delta_{\Gamma_1}^{(m,m)}
\end{bmatrix}. \\
\end{split}
\end{equation*}

We notice that the first term is the characteristic polynomial of $\Delta_{\Gamma_1 (v_1, v_1)}$. We use cofactor expansion on the $(v_2-1)^{th}$ column of the second summand to obtain the following. The labeling of this column follows from the fact that the $(v_1)^{th}$ column was previously deleted due to cofactor expansion.

\begin{equation*}
\begin{split}
X_{21}(\lambda) & = p_{\Delta_{\Gamma_1 (v_1, v_1)}}(\lambda) + \det \begin{bmatrix}
\Delta_{\Gamma_1}^{(1,1)} - \lambda & \hdots & \Delta_{\Gamma_1}^{(1,m)} \\
\vdots & \ddots & \vdots \\
\Delta_{\Gamma_1}^{(m,1)} & \hdots & \Delta_{\Gamma_1}^{(m,m)} - \lambda
\end{bmatrix} \\
& = p_{\Delta_{\Gamma_1 (v_1, v_1)}}(\lambda) + p_{\Delta_{\Gamma_1 (v_1, v_1) (v_2, v_2)}}(\lambda). \\
\end{split}
\end{equation*}

We then look at $X_{22}(\lambda)$, where we use Lemma \ref{SplitColumn_Lemma} on the $(v_2-1)^{th}$ column to obtain the following. The reason for the column labeling is the same as in the previous step.

\begin{equation*}
\begin{split}
X_{22}(\lambda) & = \det \begin{bmatrix}
\Delta_{\Gamma_1}^{(1,1)} - \lambda & \hdots & \Delta_{\Gamma_1}^{(1,v_2)} & \hdots & \Delta_{\Gamma_1}^{(1,m)} \\
\vdots & \ddots & \vdots & \ddots & \vdots \\
\Delta_{\Gamma_1}^{(v_1,1)} & \hdots & 0 & \hdots & \Delta_{\Gamma_1}^{(v_1,m)} \\
\vdots & \ddots & \vdots & \ddots & \vdots \\
\Delta_{\Gamma_1}^{(m,1)} & \hdots & \Delta_{\Gamma_1}^{(m,v_2)} & \hdots & \Delta_{\Gamma_1}^{(m,m)} - \lambda
\end{bmatrix} + \det \begin{bmatrix}
\Delta_{\Gamma_1}^{(1,1)} - \lambda & \hdots & 0 & \hdots & \Delta_{\Gamma_1}^{(1,m)} \\
\vdots & \ddots & \vdots & \ddots & \vdots \\
\Delta_{\Gamma_1}^{(v_1,1)} & \hdots & -1 & \hdots & \Delta_{\Gamma_1}^{(v_1,m)} \\
\vdots & \ddots & \vdots & \ddots & \vdots \\
\Delta_{\Gamma_1}^{(m,1)} & \hdots & 0 & \hdots & \Delta_{\Gamma_1}^{(m,m)} - \lambda
\end{bmatrix}. \\
\end{split}
\end{equation*}

We can rewrite the first summand as a minor of $\Delta_{\Gamma_1} - \lambda I$, and we use cofactor expansion on the $(v_2-1)^{th}$ column of the second term to obtain the following.

\begin{equation*}
\begin{split}
X_{22}(\lambda) & = \det \bigg( \left(\Delta_{\Gamma_1} - \lambda I\right)_{(v_2, v_1)} \bigg) + (-1) (-1)^{v_1 + (v_2-1)} \det \begin{bmatrix}
\Delta_{\Gamma_1}^{(1,1)} - \lambda & \hdots & \Delta_{\Gamma_1}^{(1,m)} \\
\vdots & \ddots & \vdots \\
\Delta_{\Gamma_1}^{(m,1)} & \hdots & \Delta_{\Gamma_1}^{(m,m)} - \lambda
\end{bmatrix} \\
& = \det \bigg( \left(\Delta_{\Gamma_1} - \lambda I\right)_{(v_2, v_1)} \bigg) + (-1) (-1)^{v_1 + (v_2-1)} p_{\Delta_{\Gamma_1 (v_1, v_1) (v_2, v_2)}}(\lambda). \\
\end{split}
\end{equation*}

Therefore, we combine $X_{21}(\lambda)$ and $X_{22}(\lambda)$ to obtain $X_2(\lambda)$,

\begin{equation*}
\begin{split}
X_2(\lambda) & = p_{\Delta_{\Gamma_1 (v_1, v_1)}}(\lambda) + p_{\Delta_{\Gamma_1 (v_1, v_1) (v_2, v_2)}}(\lambda) \\
& - (-1)^{v_2 + v_1} \left( \det\bigg( (\Delta_{\Gamma_1} - \lambda I)_{(v_2, v_1)} \bigg) - (-1)^{v_1 + (v_2-1)} p_{\Delta_{\Gamma_1 (v_1, v_1) (v_2, v_2)}}(\lambda) \right) \\
& = p_{\Delta_{\Gamma_1 (v_1, v_1)}}(\lambda) + p_{\Delta_{\Gamma_1 (v_1, v_1) (v_2, v_2)}}(\lambda) \\
& - (-1)^{v_2 + v_1} \det\bigg( (\Delta_{\Gamma_1} - \lambda I)_{(v_2, v_1)} \bigg) - p_{\Delta_{\Gamma_1 (v_1, v_1) (v_2, v_2)}}(\lambda) \\
& = p_{\Delta_{\Gamma_1 (v_1, v_1)}}(\lambda) - (-1)^{v_2 + v_1} \det\bigg( (\Delta_{\Gamma_1} - \lambda I)_{(v_2, v_1)} \bigg), \\
\end{split}
\end{equation*}

and we subsequently obtain $p_{\Delta_{\Gamma_1 \sqcup_I P_2}}(\lambda)$ by summing $X_1(\lambda)$ and $X_2(\lambda)$

\begin{equation*}
\begin{split}
p_{\Delta_{\Gamma_1 \sqcup_I P_2}}(\lambda) & = p_{\Delta_{\Gamma_1}}(\lambda) + p_{\Delta_{\Gamma_1 (v_2, v_2)}}(\lambda) - (-1)^{v_1 + v_2} \det\left(
\left(\Delta_{\Gamma_1} - \lambda I\right)_{(v_1, v_2)}
\right) \\
& + p_{\Delta_{\Gamma_1 (v_1, v_1)}}(\lambda) - (-1)^{v_2 + v_1} \det\bigg( (\Delta_{\Gamma_1} - \lambda I)_{(v_2, v_1)} \bigg).
\end{split}
\end{equation*}

Given that $(\Delta_{\Gamma_1} - \lambda I)_{(v_1, v_2)} = (\Delta_{\Gamma_1} - \lambda I)^t_{(v_2, v_1)}$ and $\det(M) = \det(M^t)$, where the $t$ denotes the transpose of the matrix, it follows that

\begin{equation*}
\begin{split}
p_{\Delta_{\Gamma_1 \sqcup_I P_2}}(\lambda) & = p_{\Delta_{\Gamma_1}}(\lambda) + p_{\Delta_{\Gamma_1 (v_1, v_1)}}(\lambda) + p_{\Delta_{\Gamma_1 (v_2, v_2)}}(\lambda) - 2 (-1)^{v_1 + v_2} \det\left(
\left(\Delta_{\Gamma_1} - \lambda I\right)_{(v_1, v_2)}
\right).
\end{split}
\end{equation*}

\end{proof}


\section{Bridge Gluing} \label{BridgeGluing_Section}

In this section, we present several properties of bridge gluing and the resulting Laplacian matrix and spectrum derived from those of the original two graphs. Our results mirror those presented in Section~\ref{InterfaceGluing_Section}, however the applications to spectral graph theory are distinct.

\begin{theorem} \label{BridgeGluingRatio_Theorem}
Let $\Gamma_1$ and $\Gamma_2$ be two graphs and $B$ a bridge graph connecting $\Gamma_1$ and $\Gamma_2$, Then, we have $\mathcal{E}(\Gamma_1 \sqcup_{B} \Gamma_2) = \frac{\mathcal{E}(\Gamma_1)\mathcal{E}(\Gamma_2)}{\mathcal{E}(B)}$.
\end{theorem}

\begin{proof}
We have
\begin{equation*}
\begin{split}
\chi(\Gamma_1 \sqcup_{B} \Gamma_2) & = |V(\Gamma_1 \sqcup_{B} \Gamma_2)| - |E(\Gamma_1 \sqcup_{B} \Gamma_2)| \\
& = (|V(\Gamma_1)| + |V(\Gamma_2)|) - (|E(\Gamma_1)| + |E(\Gamma_2)|) - |E(\Gamma_B)| \\
& = (|V(\Gamma_1)| - |E(\Gamma_1)|) + (|V(\Gamma_2)| - |E(\Gamma_2)|) - |E(\Gamma_B)|.
\end{split}
\end{equation*}
In the bridge graph $B$, we always have $\chi(B) = |V(B)| - |E(B)| = 2|E(B)|$. Hence $\chi(\Gamma_1 \sqcup_{B} \Gamma_2) = \chi(\Gamma_1) + \chi(\Gamma_2) - \chi(B)$. By Proposition~\ref{EulerCharacteristic_Proposition},
\begin{equation*}
\mathcal{E}(\Gamma_1 \sqcup_{B} \Gamma_2) = (-\lambda)^{\chi(\Gamma_1 \sqcup_{B} \Gamma_2)} = (-\lambda)^{\chi(\Gamma_1) + \chi(\Gamma_2) - \chi(B)} = \frac{(-\lambda)^{\chi(\Gamma_1)}(-\lambda)^{\chi(\Gamma_2)}}{(-\lambda)^{\chi(B)}} = \frac{\mathcal{E}(\Gamma_1)\mathcal{E}(\Gamma_2)}{\mathcal{E}(B)},
\end{equation*}
as we want.
\end{proof}


\subsection{Bridge Gluing for the Even Laplacian} \label{EvenBridgeGluing_Subsection}
In Theorem~\ref{EvenBridgeGluing_Theorem}, we derive the even Laplacian bridge gluing formula. Similar to the proof of Theorem~\ref{EvenLaplacianGluing_Theorem}, we will consider eight different cases by using the indices of the entries of the Laplacian matrix, to identify the positions of the corresponding vertices before and after bridge gluing.
\begin{theorem} \label{EvenBridgeGluing_Theorem}
Let $\Gamma_1$ and $\Gamma_2$ be two graphs, $B$ be a bridge graph, and $\Gamma = \Gamma_1 \sqcup_B \Gamma_2$. Let $\{v_1, v_2, \ldots, v_n\}$ be the vertices of $\Gamma_1$, and $\{v_{n+1}, \ldots, v_m\}$ be the vertices of $\Gamma_2$. Let $|V(B)|=q$. Then, $\Delta^{+}_{\Gamma}$ is given by
$$\Delta^+_{\Gamma}(i,j) = \begin{cases}
\Delta^{+}_{\Gamma_1}(i,j) & \text{if } i \neq j \text{ and } i,j \leq n \text{ $(1)$}\\
\Delta^{+}_{\Gamma_1}(i,i) & \text{if } i = j \text{ and } v_i \notin V(B) \text{ $(2)$}\\
\Delta^{+}_{\Gamma_1}(i,i)+1 & \text{if } i = j \text{ and } v_i \in V(B) \text{ $(3)$}\\
\Delta^{+}_{\Gamma_2}(i-n,j-n) & \text{if } i \neq j \text{ and } i,j > n \text{ $(4)$}\\
\Delta^{+}_{\Gamma_2}(i-n,i-n) & \text{if } i = j \text{ and } v_i \notin V(B) \text{ $(5)$}\\
\Delta^{+}_{\Gamma_2}(i-n,i-n)+1 & \text{if } i = j \text{ and } v_i \in V(B) \text{ $(6)$}\\
-1 & \text{if } i \leq n, j > n \text{ (or } j \leq n, i > n \text{), and } v_i,v_j \text{ are connected by a bridge} \text{ $(7)$}\\
0 & \text{otherwise.} \text{ $(8)$}
\end{cases}$$
\end{theorem}

\begin{proof}
Since bridge gluing does not create any new vertices, $|V(\Gamma)|=|V(\Gamma_1)|+|V(\Gamma_2)|$. In case (1), if $v_i$ and $v_j$ are two distinct vertices in $\Gamma_1$ (that is $i \neq j$ and $i,j \leq n$), then $\Delta^{+}_{\Gamma}(i,j)=\Delta^{+}_{\Gamma_1}(i,j)$. Similarly, in case (4), if $v_i$ and $v_j$ are two distinct vertices in $\Gamma_2$ (that is $i \neq j$ and $i,j > n$), then $\Delta^{+}_{\Gamma}(i,j)=\Delta^{+}_{\Gamma_2}(i-n,j-n)$. The reason for having $(i-n,j-n)$ is that the first $n$ vertices in $\Gamma$ belong to $\Gamma_1$.

For $v_i$, the number of its neighbors (val($v_i$)) does not change if $v_i \notin V(B)$. Thus, we have $\Delta^{+}_{\Gamma}(i,i)=\Delta^{+}_{\Gamma_1}(i,i)$ (case (2)), and $\Delta^{+}_{\Gamma}(i,i)=\Delta^{+}_{\Gamma_2}(i-n,i-n)$ (case (5)). The number of its neighbors increases by 1 if $v_i \in V(B)$, since each vertex can be connected by a bridge at most once. Thus, we have $\Delta^{+}_{\Gamma}(i,i)=\Delta^{+}_{\Gamma_1}(i,i)+1$ (case (3)), and $\Delta^{+}_{\Gamma}(i,i)=\Delta^{+}_{\Gamma_2}(i-n,i-n)+1$ (case (6)). The reason for having $(i-n,j-n)$ is the same as above.

If one of $v_i$ and $v_j$ is in $\Gamma_1$, and the other one is in $\Gamma_2$ (that is $i \leq n$ and $j>n$ or $j \leq n$ and $i>n$), then $v_i$ and $v_j$ are distinct vertices that were not connected before gluing. Thus, if $v_i$ and $v_j$ is connected by a bridge (case (7)), then $\Delta^{+}_{\Gamma}(i,i)=-1$. Otherwise (case (8)), $\Delta^{+}_{\Gamma}(i,i)=0$.
\end{proof}


\subsection{Bridge Gluing for the Odd Laplacian} \label{OddBridgeGluing_Subsection}
In Theorem~\ref{OddBridgeGluing_Theorem}, we derive the odd Laplacian bridge gluing formula. This formula gives the relationship between the odd Laplacians before and after bridge gluing, based on the positions and $\zeta$-conditions of the corresponding edges.
\begin{theorem} \label{OddBridgeGluing_Theorem}
Let $\Gamma_1$ and $\Gamma_2$ be two graphs, $B$ be the bridge graph, and $\Gamma$ is glued by $\Gamma_1$ and $\Gamma_2$ via $B$. Let $\{e_1, e_2, \ldots, \ldots, e_n\}$ be the edges of $\Gamma_1$, $\{e_{n+1}, \ldots, e_{n+q}\}$ be the bridge graph $B$, and $\{e_{n+q+1}, e_{n+q+2}, \ldots, e_m\}$ be the edges of $\Gamma_2$. Then, $\Delta^{-}_{\Gamma}$ is given by
\begin{equation*}
\Delta^{-}_{\Gamma}(i,j) = \begin{cases}
\Delta^{-}_{\Gamma_1}(i,j) & \text{if } i,j \leq n \text{ $(1)$}\\
\Delta^{-}_{\Gamma_2}(i-n-q,j-n-q) & \text{if } i,j \geq (n+q+1) \text{ $(2)$}\\
0 & \text{if } i \leq n, j \geq (n+q+1) \text{ or } j \leq n, i \geq (n+q+1) \text{ $(3)$}\\
2 & \text{if } i=j \text{ and } (n+1) \leq i \leq (n+q) \text{ $(4)$}\\ 
0 & \text{if } i \neq j \text{ and } (n+1) \leq i,j \leq (n+q) \text{ $(5)$}\\
\zeta & \text{otherwise,} \text{ $(6)$}
\end{cases}
\end{equation*}
where 
\begin{equation*}
\zeta=\begin{cases}
0 & \text{if } e_i \text{ is not incident to } e_j\\
1 & \text{if } e_i \text{ is incident to } e_j \text{ at } v_s \text{, and } e_i \text{ and } e_j \text{ both start or end at } v_s\\
-1 & \text{if } e_i \text{ is incident to } e_j \text{ at } v_s \text{, and one of the edges starts at } v_s \text{, the other ends at } v_s. 
\end{cases}
\end{equation*}
\end{theorem}

\begin{proof}
Since bridge gluing does not change the $\zeta$-condition and orientations nor eliminate any edge in both $\Gamma_1$ and $\Gamma_2$, we have $\Delta^{-}_{\Gamma}(i,j)=\Delta^{-}_{\Gamma_1}(i,j)$ if both $e_i$ and $e_j$ are in $E(\Gamma_1)$ (case (1)), or $\Delta^{-}_{\Gamma}(i,j)=\Delta^{-}_{\Gamma_2}(i-n-q,j-n-q)$ if both $e_i$ and $e_j$ are in $E(\Gamma_2)$ (case (2)), by Proposition~\ref{OddLaplacian_Proposition}. The reason for having $(i-n-q,j-n-q)$ is that the first $n+q$ edges in $\Gamma$ belong to either $\Gamma_1$ or $B$.

When one of $e_i$ and $e_j$ is in $E(\Gamma_1)$ and the other is in $E(\Gamma_2)$, that is $i \leq n, j \geq (n+q+1)$ or $j \leq n, i \geq (n+q+1)$ (case (3)), $e_i$ and $e_j$ are not incident before and after gluing. Thus, we have $\Delta^{-}_{\Gamma}(i,j)=0$.

When $(n+1) \leq i,j \leq (n+q)$, we have both $e_i$ and $e_j$ in $E(B)$. Therefore, if $i=j$ (case (4)), we have $\Delta^{-}_{\Gamma}(i,i)=2$, and if $i \neq j$ (case (5)), we have $\Delta^{-}_{\Gamma}(i,j)=0$, by Proposition~\ref{OddLaplacian_Proposition}. When exactly one of $(n+1) \leq i \leq (n+q)$ and $(n+1) \leq j \leq (n+q)$ is true (case (6)), we have exactly one of $e_i$ and $e_j$ is in $E(B)$. Hence, by Proposition~\ref{OddLaplacian_Proposition}, we get the desired value of $\zeta$ for all three $\zeta$-conditions between $e_i$ and $e_j$.
\end{proof}


\subsection{Bridge Gluing Spectra} \label{BridgeGluingSpectra_Subsection}

An explicit formula for the spectrum of the Laplacian after 1-bridge gluing, i.e. $|E(B)| = 1$, is presented. This outlines the basis for iteratively calculating the spectrum when $|E(B)| > 1$, which is introduced in Section \ref{Algorithm_Subsection}.

\begin{theorem} \label{EdgeBridgeGluingSpectrum_Theorem}
Let $\Gamma_1$ and $\Gamma_2$ be two graphs with $m$ and $n$ vertices respectively, and identify a bridge graph $B$ with $V(B) = \{v^1_i, v^2_i\}$ and $E(B) = \{e_i\}$ where $v^1_i \in V(\Gamma_1)$, $v^2_i \in V(\Gamma_2)$, and the endpoints of $e_i$ are $v^1_i$ and $v^2_i$. Then,
\begin{equation*}
\begin{split}
p_{\Delta_{\Gamma_1 \sqcup_B \Gamma_2}}(\lambda) & = p_{\Delta_{\Gamma_1}}(\lambda) p_{\Delta_{\Gamma_2}}(\lambda)+ p_{\Delta_{\Gamma_1}}(\lambda) p_{\Delta_{\Gamma_{2 \left( v^2_i, v^2_i \right)}}}(\lambda) + p_{\Delta_{\Gamma_{1 \left( v^1_i, v^1_i \right)}}}(\lambda) p_{\Delta_{\Gamma_2}}(\lambda).
\end{split}
\end{equation*}
\end{theorem}

\begin{proof}
From Definition~\ref{EvenGraphLaplacian_Definition}, we see that $\Delta_{\Gamma_1 \sqcup_B \Gamma_2}$ is the following matrix: \footnote{Since the matrices we consider are large, we denote $M^{(i,j)}$ as the $(i,j)$-entry of matrix $M$ in this proof.}

\begin{equation*}
\begin{bmatrix} 
\Delta^{(1,1)}_{\Gamma_1} & \hdots & \Delta^{(1,v^1_i)}_{\Gamma_1} & \hdots & \Delta^{(1,m)}_{\Gamma_1} & 0 & \hdots & 0 & \hdots & 0 \\
\vdots & \ddots & \vdots & \ddots & \vdots & \vdots & \ddots & \vdots & \ddots & \vdots \\
\Delta^{(v^1_i,1)}_{\Gamma_1} & \hdots & \Delta^{(v^1_i, v^1_i)}_{\Gamma_1} + 1 & \hdots & \Delta^{(v^1_i,m)}_{\Gamma_1} & 0 & \hdots & -1 & \hdots & 0 \\
\vdots & \ddots & \vdots & \ddots & \vdots & \vdots & \ddots & \vdots & \ddots & \vdots \\
\Delta^{(m,1)}_{\Gamma_1} & \hdots & \Delta^{(m, v^1_i)}_{\Gamma_1} & \hdots & \Delta^{(m,m)}_{\Gamma_1}& 0 & \hdots & 0 & \hdots & 0 \\
0 & \hdots & 0 & \hdots & 0 & \Delta^{(1,1)}_{\Gamma_2} - \lambda & \hdots & \Delta^{(1,v^2_i)}_{\Gamma_2} & \hdots & \Delta^{(1,n)}_{\Gamma_2} \\
\vdots & \ddots & \vdots & \ddots & \vdots & \vdots & \ddots & \vdots & \ddots & \vdots \\
0 & \hdots & -1 & \hdots & 0 & \Delta^{(v^2_i,1)}_{\Gamma_2} & \hdots & \Delta^{(v^2_i,v^2_i)}_{\Gamma_2} + 1 & \hdots & \Delta^{(v^2_i,n)}_{\Gamma_2} \\
\vdots & \ddots & \vdots & \ddots & \vdots & \vdots & \ddots & \vdots & \ddots & \vdots \\
0 & \hdots & 0 & \hdots & 0 & \Delta^{(n,1)}_{\Gamma_2} & \hdots & \Delta^{(n,v^2_i)}_{\Gamma_2} & \hdots & \Delta^{(n,n)}_{\Gamma_2} \\
\end{bmatrix}. \\
\end{equation*}

We know that the characteristic polynomial of $\Delta_{\Gamma_1 \sqcup_B \Gamma_2}$ is then given by

\begin{equation*} 
\det \begin{bmatrix} 
\Delta^{(1,1)}_{\Gamma_1} - \lambda & \hdots & \Delta^{(1,v^1_i)}_{\Gamma_1} & \hdots & \Delta^{(1,m)}_{\Gamma_1} & 0 & \hdots & 0 & \hdots & 0 \\
\vdots & \ddots & \vdots & \ddots & \vdots & \vdots & \ddots & \vdots & \ddots & \vdots \\
\Delta^{(v^1_i,1)}_{\Gamma_1} & \hdots & \Delta^{(v^1_i, v^1_i)}_{\Gamma_1} + 1 - \lambda & \hdots & \Delta^{(v^1_i,m)}_{\Gamma_1} & 0 & \hdots & -1 & \hdots & 0 \\
\vdots & \ddots & \vdots & \ddots & \vdots & \vdots & \ddots & \vdots & \ddots & \vdots \\
\Delta^{(m,1)}_{\Gamma_1} & \hdots & \Delta^{(m, v^1_i)}_{\Gamma_1} & \hdots & \Delta^{(m,m)}_{\Gamma_1} - \lambda & 0 & \hdots & 0 & \hdots & 0 \\
0 & \hdots & 0 & \hdots & 0 & \Delta^{(1,1)}_{\Gamma_2} - \lambda & \hdots & \Delta^{(1,v^2_i)}_{\Gamma_2} & \hdots & \Delta^{(1,n)}_{\Gamma_2} \\
\vdots & \ddots & \vdots & \ddots & \vdots & \vdots & \ddots & \vdots & \ddots & \vdots \\
0 & \hdots & -1 & \hdots & 0 & \Delta^{(v^2_i,1)}_{\Gamma_2} & \hdots & \Delta^{(v^2_i,v^2_i)}_{\Gamma_2} + 1 - \lambda & \hdots & \Delta^{(v^2_i,n)}_{\Gamma_2} \\
\vdots & \ddots & \vdots & \ddots & \vdots & \vdots & \ddots & \vdots & \ddots & \vdots \\
0 & \hdots & 0 & \hdots & 0 & \Delta^{(n,1)}_{\Gamma_2} & \hdots & \Delta^{(n,v^2_i)}_{\Gamma_2} & \hdots & \Delta^{(n,n)}_{\Gamma_2} - \lambda \\
\end{bmatrix}. \\
\end{equation*}

By Lemma~\ref{SplitColumn_Lemma}, we can decompose $p_{\Delta_{\Gamma_1 \sqcup_B \Gamma_2}}(\lambda)$ into the following sum:

\begin{equation*}
\det \begin{bmatrix} 
\Delta^{(1,1)}_{\Gamma_1} - \lambda & \hdots & \Delta^{(1,v^1_i)}_{\Gamma_1} & \hdots & \Delta^{(1,m)}_{\Gamma_1} & 0 & \hdots & 0 & \hdots & 0 \\
\vdots & \ddots & \vdots & \ddots & \vdots & \vdots & \ddots & \vdots & \ddots & \vdots \\
\Delta^{(v^1_i,1)}_{\Gamma_1} & \hdots & \Delta^{(v^1_i, v^1_i)}_{\Gamma_1} - \lambda & \hdots & \Delta^{(v^1_i,m)}_{\Gamma_1} & 0 & \hdots & -1 & \hdots & 0 \\
\vdots & \ddots & \vdots & \ddots & \vdots & \vdots & \ddots & \vdots & \ddots & \vdots \\
\Delta^{(m,1)}_{\Gamma_1} & \hdots & \Delta^{(m, v^1_i)}_{\Gamma_1} & \hdots & \Delta^{(m,m)}_{\Gamma_1} - \lambda & 0 & \hdots & 0 & \hdots & 0 \\
0 & \hdots & 0 & \hdots & 0 & \Delta^{(1,1)}_{\Gamma_2} - \lambda & \hdots & \Delta^{(1,v^2_i)}_{\Gamma_2} & \hdots & \Delta^{(1,n)}_{\Gamma_2} \\
\vdots & \ddots & \vdots & \ddots & \vdots & \vdots & \ddots & \vdots & \ddots & \vdots \\
0 & \hdots & 0 & \hdots & 0 & \Delta^{(v^2_i,1)}_{\Gamma_2} & \hdots & \Delta^{(v^2_i,v^2_i)}_{\Gamma_2} + 1 - \lambda & \hdots & \Delta^{(v^2_i,n)}_{\Gamma_2} \\
\vdots & \ddots & \vdots & \ddots & \vdots & \vdots & \ddots & \vdots & \ddots & \vdots \\
0 & \hdots & 0 & \hdots & 0 & \Delta^{(n,1)}_{\Gamma_2} & \hdots & \Delta^{(n,v^2_i)}_{\Gamma_2} & \hdots & \Delta^{(n,n)}_{\Gamma_2} - \lambda \\
\end{bmatrix}
\end{equation*}
\begin{equation*}
+\det \begin{bmatrix} 
\Delta^{(1,1)}_{\Gamma_1} - \lambda & \hdots & 0 & \hdots & \Delta^{(1,m)}_{\Gamma_1} & 0 & \hdots & 0 & \hdots & 0 \\
\vdots & \ddots & \vdots & \ddots & \vdots & \vdots & \ddots & \vdots & \ddots & \vdots \\
\Delta^{(v^1_i,1)}_{\Gamma_1} & \hdots & 1 & \hdots & \Delta^{(v^1_i,m)}_{\Gamma_1} & 0 & \hdots & -1 & \hdots & 0 \\
\vdots & \ddots & \vdots & \ddots & \vdots & \vdots & \ddots & \vdots & \ddots & \vdots \\
\Delta^{(m,1)}_{\Gamma_1} & \hdots & 0 & \hdots & \Delta^{(m,m)}_{\Gamma_1} - \lambda & 0 & \hdots & 0 & \hdots & 0 \\
0 & \hdots & 0 & \hdots & 0 & \Delta^{(1,1)}_{\Gamma_2} - \lambda & \hdots & \Delta^{(1,v^2_i)}_{\Gamma_2} & \hdots & \Delta^{(1,n)}_{\Gamma_2} \\
\vdots & \ddots & \vdots & \ddots & \vdots & \vdots & \ddots & \vdots & \ddots & \vdots \\
0 & \hdots & -1 & \hdots & 0 & \Delta^{(v^2_i,1)}_{\Gamma_2} & \hdots & \Delta^{(v^2_i,v^2_i)}_{\Gamma_2} + 1 - \lambda & \hdots & \Delta^{(v^2_i,n)}_{\Gamma_2} \\
\vdots & \ddots & \vdots & \ddots & \vdots & \vdots & \ddots & \vdots & \ddots & \vdots \\
0 & \hdots & 0 & \hdots & 0 & \Delta^{(n,1)}_{\Gamma_2} & \hdots & \Delta^{(n,v^2_i)}_{\Gamma_2} & \hdots & \Delta^{(n,n)}_{\Gamma_2} - \lambda \\
\end{bmatrix}.
\end{equation*}

The first summand is a block upper triangular matrix, so we can rewrite it using Lemma~\ref{BlockTriangular_Lemma}. Additionally, we can cofactor expand the second determinant along the $(v^1_i)$-column to obtain the following:

\begin{equation*}
\begin{split}
p_{\Delta_{\Gamma_1 \sqcup_B \Gamma_2}}(\lambda) & = \det \begin{bmatrix} 
\Delta^{(1,1)}_{\Gamma_1} - \lambda & \hdots & \Delta^{(1,v^1_i)}_{\Gamma_1} & \hdots & \Delta^{(1,m)}_{\Gamma_1} \\
\vdots & \ddots & \vdots & \ddots & \vdots \\
\Delta^{(v^1_i,1)}_{\Gamma_1} & \hdots & \Delta^{(v^1_i, v^1_i)}_{\Gamma_1} - \lambda & \hdots & \Delta^{(v^1_i,m)}_{\Gamma_1} \\
\vdots & \ddots & \vdots & \ddots & \vdots \\
\Delta^{(m,1)}_{\Gamma_1} & \hdots & \Delta^{(m, v^1_i)}_{\Gamma_1} & \hdots & \Delta^{(m,m)}_{\Gamma_1} - \lambda \\
\end{bmatrix} \\
& \cdot \det \begin{bmatrix}
\Delta^{(1,1)}_{\Gamma_2} - \lambda & \hdots & \Delta^{(1,v^2_i)}_{\Gamma_2} & \hdots & \Delta^{(1,n)}_{\Gamma_2} \\
\vdots & \ddots & \vdots & \ddots & \vdots \\
\Delta^{(v^2_i,1)}_{\Gamma_2} & \hdots & \Delta^{(v^2_i,v^2_i)}_{\Gamma_2} + 1 - \lambda & \hdots & \Delta^{(v^2_i,n)}_{\Gamma_2} \\
\vdots & \ddots & \vdots & \ddots & \vdots \\
\Delta^{(n,1)}_{\Gamma_2} & \hdots & \Delta^{(n,v^2_i)}_{\Gamma_2} & \hdots & \Delta^{(n,n)}_{\Gamma_2} - \lambda \\
\end{bmatrix} \\
& + \det \begin{bmatrix} 
\Delta^{(1,1)}_{\Gamma_1} - \lambda & \hdots & \Delta^{(1,m)}_{\Gamma_1} & 0 & \hdots & 0 & \hdots & 0 \\
\vdots & \ddots & \vdots & \vdots & \ddots & \vdots & \ddots & \vdots \\
\Delta^{(m,1)}_{\Gamma_1} & \hdots & \Delta^{(m,m)}_{\Gamma_1} - \lambda & 0 & \hdots & 0 & \hdots & 0 \\
0 & \hdots & 0 & \Delta^{(1,1)}_{\Gamma_2} - \lambda & \hdots & \Delta^{(1,v^2_i)}_{\Gamma_2} & \hdots & \Delta^{(1,n)}_{\Gamma_2} \\
\vdots & \ddots & \vdots & \vdots & \ddots & \vdots & \ddots & \vdots \\
0 & \hdots & 0 & \Delta^{(v^2_i,1)}_{\Gamma_2} & \hdots & \Delta^{(v^2_i,v^2_i)}_{\Gamma_2} + 1 - \lambda & \hdots & \Delta^{(v^2_i,n)}_{\Gamma_2} \\
\vdots & \ddots & \vdots & \vdots & \ddots & \vdots & \ddots & \vdots \\
0 & \hdots & 0 & \Delta^{(n,1)}_{\Gamma_2} & \hdots & \Delta^{(n,v^2_i)}_{\Gamma_2} & \hdots & \Delta^{(n,n)}_{\Gamma_2} - \lambda \\
\end{bmatrix} \\
& + (-1) (-1)^{(m+v^2_i) + v^1_i} \det \begin{bmatrix} 
\Delta^{(1,1)}_{\Gamma_1} - \lambda & \hdots & \Delta^{(1,m)}_{\Gamma_1} & 0 & \hdots & 0 & \hdots & 0 \\
\vdots & \ddots & \vdots & \vdots & \ddots & \vdots & \ddots & \vdots \\
\Delta^{(v^1_i,1)}_{\Gamma_1} & \hdots & \Delta^{(v^1_i,m)}_{\Gamma_1} & 0 & \hdots & -1 & \hdots & 0 \\
\vdots & \ddots & \vdots & \vdots & \ddots & \vdots & \ddots & \vdots \\
\Delta^{(m,1)}_{\Gamma_1} & \hdots & \Delta^{(m,m)}_{\Gamma_1} - \lambda & 0 & \hdots & 0 & \hdots & 0 \\
0 & \hdots & 0 & \Delta^{(1,1)}_{\Gamma_2} - \lambda & \hdots & \Delta^{(1,v^2_i)}_{\Gamma_2} & \hdots & \Delta^{(1,n)}_{\Gamma_2} \\
\vdots & \ddots & \vdots & \vdots & \ddots & \vdots & \ddots & \vdots \\
0 & \hdots & 0 & \Delta^{(n,1)}_{\Gamma_2} & \hdots & \Delta^{(n,v^2_i)}_{\Gamma_2} & \hdots & \Delta^{(n,n)}_{\Gamma_2} - \lambda \\
\end{bmatrix} \\
& =: p_{\Delta_{\Gamma_1}}(\lambda) Y_1(\lambda) + Y_2(\lambda) + (-1)(-1)^{(m+v^2_i)+v^1_i} Y_3(\lambda).
\end{split}
\end{equation*}

We first decompose $Y_1(\lambda)$ using Lemma~\ref{SplitColumn_Lemma} and cofactor expansion.

\begin{equation*}
\begin{split}
Y_1(\lambda) & = \det \begin{bmatrix}
\Delta^{(1,1)}_{\Gamma_2} - \lambda & \hdots & \Delta^{(1,v^2_i)}_{\Gamma_2} & \hdots & \Delta^{(1,n)}_{\Gamma_2} \\
\vdots & \ddots & \vdots & \ddots & \vdots \\
\Delta^{(v^2_i,1)}_{\Gamma_2} & \hdots & \Delta^{(v^2_i,v^2_i)}_{\Gamma_2} - \lambda & \hdots & \Delta^{(v^2_i,n)}_{\Gamma_2} \\
\vdots & \ddots & \vdots & \ddots & \vdots \\
\Delta^{(n,1)}_{\Gamma_2} & \hdots & \Delta^{(n,v^2_i)}_{\Gamma_2} & \hdots & \Delta^{(n,n)}_{\Gamma_2} - \lambda \\
\end{bmatrix} + \det \begin{bmatrix}
\Delta^{(1,1)}_{\Gamma_2} - \lambda & \hdots & 0 & \hdots & \Delta^{(1,n)}_{\Gamma_2} \\
\vdots & \ddots & \vdots & \ddots & \vdots \\
\Delta^{(v^2_i,1)}_{\Gamma_2} & \hdots & 1 & \hdots & \Delta^{(v^2_i,n)}_{\Gamma_2} \\
\vdots & \ddots & \vdots & \ddots & \vdots \\
\Delta^{(n,1)}_{\Gamma_2} & \hdots & 0 & \hdots & \Delta^{(n,n)}_{\Gamma_2} - \lambda \\
\end{bmatrix} \\
& = p_{\Delta_{\Gamma_2}}(\lambda) + \det \begin{bmatrix}
\Delta^{(1,1)}_{\Gamma_2} - \lambda & \hdots & 0 & \hdots & \Delta^{(1,n)}_{\Gamma_2} \\
\vdots & \ddots & \vdots & \ddots & \vdots \\
\Delta^{(v^2_i,1)}_{\Gamma_2} & \hdots & 1 & \hdots & \Delta^{(v^2_i,n)}_{\Gamma_2} \\
\vdots & \ddots & \vdots & \ddots & \vdots \\
\Delta^{(n,1)}_{\Gamma_2} & \hdots & 0 & \hdots & \Delta^{(n,n)}_{\Gamma_2} - \lambda \\
\end{bmatrix}. \\
\end{split}
\end{equation*}

We perform cofactor expansion on the second summand along the $(v^2_i)$-column, where we notice that the $1$ is located along the diagonal of this matrix.

\begin{equation*}
\begin{split}
Y_1(\lambda) & = p_{\Delta_{\Gamma_2}}(\lambda) + \det \begin{bmatrix}
\Delta^{(1,1)}_{\Gamma_2} - \lambda & \hdots & \Delta^{(1,n)}_{\Gamma_2} \\
\vdots & \ddots & \vdots \\
\Delta^{(n,1)}_{\Gamma_2} & \hdots & \Delta^{(n,n)}_{\Gamma_2} - \lambda \\
\end{bmatrix} = p_{\Delta_{\Gamma_2}}(\lambda) + p_{\Delta_{\Gamma_2 \left(v^2_i, v^2_i\right)}}(\lambda).
\end{split}
\end{equation*}

We now split $Y_2(\lambda)$ using Lemma~\ref{SplitColumn_Lemma}.

\begin{equation*}
\begin{split}
Y_2(\lambda) & = \det \begin{bmatrix} 
\Delta^{(1,1)}_{\Gamma_1} - \lambda & \hdots & \Delta^{(1,m)}_{\Gamma_1} & 0 & \hdots & 0 & \hdots & 0 \\
\vdots & \ddots & \vdots & \vdots & \ddots & \vdots & \ddots & \vdots \\
\Delta^{(m,1)}_{\Gamma_1} & \hdots & \Delta^{(m,m)}_{\Gamma_1} - \lambda & 0 & \hdots & 0 & \hdots & 0 \\
0 & \hdots & 0 & \Delta^{(1,1)}_{\Gamma_2} - \lambda & \hdots & \Delta^{(1,v^2_i)}_{\Gamma_2} & \hdots & \Delta^{(1,n)}_{\Gamma_2} \\
\vdots & \ddots & \vdots & \vdots & \ddots & \vdots & \ddots & \vdots \\
0 & \hdots & 0 & \Delta^{(v^2_i,1)}_{\Gamma_2} & \hdots & \Delta^{(v^2_i,v^2_i)}_{\Gamma_2} - \lambda & \hdots & \Delta^{(v^2_i,n)}_{\Gamma_2} \\
\vdots & \ddots & \vdots & \vdots & \ddots & \vdots & \ddots & \vdots \\
0 & \hdots & 0 & \Delta^{(n,1)}_{\Gamma_2} & \hdots & \Delta^{(n,v^2_i)}_{\Gamma_2} & \hdots & \Delta^{(n,n)}_{\Gamma_2} - \lambda \\
\end{bmatrix} \\
& + \det \begin{bmatrix} 
\Delta^{(1,1)}_{\Gamma_1} - \lambda & \hdots & \Delta^{(1,m)}_{\Gamma_1} & 0 & \hdots & 0 & \hdots & 0 \\
\vdots & \ddots & \vdots & \vdots & \ddots & \vdots & \ddots & \vdots \\
\Delta^{(m,1)}_{\Gamma_1} & \hdots & \Delta^{(m,m)}_{\Gamma_1} - \lambda & 0 & \hdots & 0 & \hdots & 0 \\
0 & \hdots & 0 & \Delta^{(1,1)}_{\Gamma_2} - \lambda & \hdots & 0 & \hdots & \Delta^{(1,n)}_{\Gamma_2} \\
\vdots & \ddots & \vdots & \vdots & \ddots & \vdots & \ddots & \vdots \\
0 & \hdots & 0 & \Delta^{(v^2_i,1)}_{\Gamma_2} & \hdots & 1 & \hdots & \Delta^{(v^2_i,n)}_{\Gamma_2} \\
\vdots & \ddots & \vdots & \vdots & \ddots & \vdots & \ddots & \vdots \\
0 & \hdots & 0 & \Delta^{(n,1)}_{\Gamma_2} & \hdots & 0 & \hdots & \Delta^{(n,n)}_{\Gamma_2} - \lambda \\
\end{bmatrix}. \\
\end{split}
\end{equation*}

We notice that both summands are block matrices, so by Lemma~\ref{BlockTriangular_Lemma}, we have the following.

\begin{equation*}
\begin{split}
Y_2(\lambda) & = \det \begin{bmatrix} 
\Delta^{(1,1)}_{\Gamma_1} - \lambda & \hdots & \Delta^{(1,m)}_{\Gamma_1} \\
\vdots & \ddots & \vdots \\
\Delta^{(m,1)}_{\Gamma_1} & \hdots & \Delta^{(m,m)}_{\Gamma_1} - \lambda \\
\end{bmatrix}
\det \begin{bmatrix} 
\Delta^{(1,1)}_{\Gamma_2} - \lambda & \hdots & \Delta^{(1,v^2_i)}_{\Gamma_2} & \hdots & \Delta^{(1,n)}_{\Gamma_2} \\
\vdots & \ddots & \vdots & \ddots & \vdots \\
\Delta^{(v^2_i,1)}_{\Gamma_2} & \hdots & \Delta^{(v^2_i,v^2_i)}_{\Gamma_2} - \lambda & \hdots & \Delta^{(v^2_i,n)}_{\Gamma_2} \\
\vdots & \ddots & \vdots & \ddots & \vdots \\
\Delta^{(n,1)}_{\Gamma_2} & \hdots & \Delta^{(n,v^2_i)}_{\Gamma_2} & \hdots & \Delta^{(n,n)}_{\Gamma_2} - \lambda \\
\end{bmatrix} \\
& + \det \begin{bmatrix} 
\Delta^{(1,1)}_{\Gamma_1} - \lambda & \hdots & \Delta^{(1,m)}_{\Gamma_1} \\
\vdots & \ddots & \vdots \\
\Delta^{(m,1)}_{\Gamma_1} & \hdots & \Delta^{(m,m)}_{\Gamma_1} - \lambda \\
\end{bmatrix}
\det \begin{bmatrix} 
\Delta^{(1,1)}_{\Gamma_2} - \lambda & \hdots & 0 & \hdots & \Delta^{(1,n)}_{\Gamma_2} \\
\vdots & \ddots & \vdots & \ddots & \vdots \\
\Delta^{(v^2_i,1)}_{\Gamma_2} & \hdots & 1 & \hdots & \Delta^{(v^2_i,n)}_{\Gamma_2} \\
\vdots & \ddots & \vdots & \ddots & \vdots \\
\Delta^{(n,1)}_{\Gamma_2} & \hdots & 0 & \hdots & \Delta^{(n,n)}_{\Gamma_2} - \lambda \\
\end{bmatrix} \\
& = p_{\Delta_{\Gamma_1 \left(v^1_i, v^1_i\right)}}(\lambda)
p_{\Delta_{\Gamma_2}}(\lambda) + p_{\Delta_{\Gamma_1 \left(v^1_i, v^1_i\right)}}(\lambda)
\det \begin{bmatrix} 
\Delta^{(1,1)}_{\Gamma_2} - \lambda & \hdots & 0 & \hdots & \Delta^{(1,n)}_{\Gamma_2} \\
\vdots & \ddots & \vdots & \ddots & \vdots \\
\Delta^{(v^2_i,1)}_{\Gamma_2} & \hdots & 1 & \hdots & \Delta^{(v^2_i,n)}_{\Gamma_2} \\
\vdots & \ddots & \vdots & \ddots & \vdots \\
\Delta^{(n,1)}_{\Gamma_2} & \hdots & 0 & \hdots & \Delta^{(n,n)}_{\Gamma_2} - \lambda \\
\end{bmatrix}. \\
\end{split}
\end{equation*}

The last determinant can be obtained by cofactor expansion along the $(v^2_i)$-column.

\begin{equation*}
\begin{split}
Y_2(\lambda) & = p_{\Delta_{\Gamma_1 \left(v^1_i, v^1_i\right)}}(\lambda)
p_{\Delta_{\Gamma_2}}(\lambda) + p_{\Delta_{\Gamma_1 \left(v^1_i, v^1_i\right)}}(\lambda)
\det \begin{bmatrix} 
\Delta^{(1,1)}_{\Gamma_2} - \lambda & \hdots & \Delta^{(1,n)}_{\Gamma_2} \\
\vdots & \ddots & \vdots \\
\Delta^{(n,1)}_{\Gamma_2} & \hdots & \Delta^{(n,n)}_{\Gamma_2} - \lambda \\
\end{bmatrix} \\
& = p_{\Delta_{\Gamma_1 \left(v^1_i, v^1_i\right)}}(\lambda)
p_{\Delta_{\Gamma_2}}(\lambda) + p_{\Delta_{\Gamma_1 \left(v^1_i, v^1_i\right)}}(\lambda) p_{\Delta_{\Gamma_2 \left(v^2_i, v^2_i\right)}}(\lambda).
\end{split}
\end{equation*}

We lastly split $Y_3(\lambda)$ using Lemma~\ref{SplitColumn_Lemma}.

\begin{equation*}
\begin{split}
Y_3(\lambda) & = \det \begin{bmatrix} 
\Delta^{(1,1)}_{\Gamma_1} - \lambda & \hdots & \Delta^{(1,m)}_{\Gamma_1} & 0 & \hdots & 0 & \hdots & 0 \\
\vdots & \ddots & \vdots & \vdots & \ddots & \vdots & \ddots & \vdots \\
\Delta^{(v^1_i,1)}_{\Gamma_1} & \hdots & \Delta^{(v^1_i,m)}_{\Gamma_1} & 0 & \hdots & 0 & \hdots & 0 \\
\vdots & \ddots & \vdots & \vdots & \ddots & \vdots & \ddots & \vdots \\
\Delta^{(m,1)}_{\Gamma_1} & \hdots & \Delta^{(m,m)}_{\Gamma_1} - \lambda & 0 & \hdots & 0 & \hdots & 0 \\
0 & \hdots & 0 & \Delta^{(1,1)}_{\Gamma_2} - \lambda & \hdots & \Delta^{(1,v^2_i)}_{\Gamma_2} & \hdots & \Delta^{(1,n)}_{\Gamma_2} \\
\vdots & \ddots & \vdots & \vdots & \ddots & \vdots & \ddots & \vdots \\
0 & \hdots & 0 & \Delta^{(n,1)}_{\Gamma_2} & \hdots & \Delta^{(n,v^2_i)}_{\Gamma_2} & \hdots & \Delta^{(n,n)}_{\Gamma_2} - \lambda \\
\end{bmatrix} \\
& + \det \begin{bmatrix} 
\Delta^{(1,1)}_{\Gamma_1} - \lambda & \hdots & \Delta^{(1,m)}_{\Gamma_1} & 0 & \hdots & 0 & \hdots & 0 \\
\vdots & \ddots & \vdots & \vdots & \ddots & \vdots & \ddots & \vdots \\
\Delta^{(v^1_i,1)}_{\Gamma_1} & \hdots & \Delta^{(v^1_i,m)}_{\Gamma_1} & 0 & \hdots & -1 & \hdots & 0 \\
\vdots & \ddots & \vdots & \vdots & \ddots & \vdots & \ddots & \vdots \\
\Delta^{(m,1)}_{\Gamma_1} & \hdots & \Delta^{(m,m)}_{\Gamma_1} - \lambda & 0 & \hdots & 0 & \hdots & 0 \\
0 & \hdots & 0 & \Delta^{(1,1)}_{\Gamma_2} - \lambda & \hdots & 0 & \hdots & \Delta^{(1,n)}_{\Gamma_2} \\
\vdots & \ddots & \vdots & \vdots & \ddots & \vdots & \ddots & \vdots \\
0 & \hdots & 0 & \Delta^{(n,1)}_{\Gamma_2} & \hdots & 0 & \hdots & \Delta^{(n,n)}_{\Gamma_2} - \lambda \\
\end{bmatrix}. \\
\end{split}
\end{equation*}

We notice that the first summand is a block lower triangular matrix, where the upper left block has a $0$-column. Therefore, the first determinant is $0$, leaving the second determinant to be cofactor expanded along the $(m-1+v^2_i)$-column. The $-1$ comes from the fact that a column has previously been deleted for cofactor expansion.

\begin{equation*}
\begin{split}
Y_3(\lambda) & = (-1) (-1)^{v^1_i + (m-1+v^2_i)} \det \begin{bmatrix} 
\Delta^{(1,1)}_{\Gamma_1} - \lambda & \hdots & \Delta^{(1,m)}_{\Gamma_1} & 0 & \hdots & 0 \\
\vdots & \ddots & \vdots & \vdots & \ddots & \vdots \\
\Delta^{(m,1)}_{\Gamma_1} & \hdots & \Delta^{(m,m)}_{\Gamma_1} - \lambda & 0 & \hdots & 0 \\
0 & \hdots & 0 & \Delta^{(1,1)}_{\Gamma_2} - \lambda & \hdots & \Delta^{(1,n)}_{\Gamma_2} \\
\vdots & \ddots & \vdots & \vdots & \ddots & \vdots \\
0 & \hdots & 0 & \Delta^{(n,1)}_{\Gamma_2} & \hdots & \Delta^{(n,n)}_{\Gamma_2} - \lambda \\
\end{bmatrix} \\
& = (-1) (-1)^{v^1_i + (m-1+v^2_i)} p_{\Delta_{\Gamma_1 \left(v^1_i, v^1_i\right)}}(\lambda) p_{\Delta_{\Gamma_2 \left(v^2_i, v^2_i\right)}}(\lambda).
\end{split}
\end{equation*}

By putting everything together, we obtain the following expression for $p_{\Delta_{\Gamma_1 \sqcup_B \Gamma_2}}(\lambda)$:

\begin{equation*}
\begin{split}
p_{\Delta_{\Gamma_1 \sqcup_B \Gamma_2}}(\lambda) & = p_{\Delta_{\Gamma_1}}(\lambda) Y_1(\lambda) + Y_2(\lambda) + (-1)(-1)^{(m+v^2_i)+v^1_i} Y_3(\lambda) \\
& = p_{\Delta_{\Gamma_1}}(\lambda) \left( p_{\Delta_{\Gamma_2}}(\lambda) + p_{\Delta_{\Gamma_2 \left(v^2_i, v^2_i\right)}}(\lambda) \right) + p_{\Delta_{\Gamma_1 \left(v^1_i, v^1_i\right)}}(\lambda) p_{\Delta_{\Gamma_2}}(\lambda) + p_{\Delta_{\Gamma_1 \left(v^1_i, v^1_i\right)}}(\lambda) p_{\Delta_{\Gamma_2 \left(v^2_i, v^2_i\right)}}(\lambda) \\
& + (-1)(-1)^{(m+v^2_i)+v^1_i} (-1) (-1)^{v^1_i + (m-1+v^2_i)} p_{\Delta_{\Gamma_1 \left(v^1_i, v^1_i\right)}}(\lambda) p_{\Delta_{\Gamma_2 \left(v^2_i, v^2_i\right)}}(\lambda) \\
& = p_{\Delta_{\Gamma_1}}(\lambda) p_{\Delta_{\Gamma_2}}(\lambda) + p_{\Delta_{\Gamma_1}}(\lambda) p_{\Delta_{\Gamma_2 \left(v^2_i, v^2_i\right)}}(\lambda) + p_{\Delta_{\Gamma_1 \left(v^1_i, v^1_i\right)}}(\lambda) p_{\Delta_{\Gamma_2}}(\lambda).
\end{split}
\end{equation*}

\end{proof}


\subsection{Algorithm for Computing the Spectrum of the General Bridge Gluing} \label{Algorithm_Subsection}

The purpose here is to calculate the spectrum of $\Gamma_1 \sqcup_B \Gamma_2$, where $\Gamma_1$ and $\Gamma_2$ are initially disjoint, $\{v^1_1, \ldots, v^1_k\} \in V(\Gamma_1)$, $\{v^2_1, \ldots, v^2_k\} \in V(\Gamma_2)$, and $E(B) = \{e_1, \ldots, e_k\}$ where all $e_i \in E(B)$ have distinct endpoints $(v^1_1, v^2_1), \ldots, (v^1_k, v^2_k)$. Using Theorems~\ref{EdgeInterfaceSpectrum_Theorem} and~\ref{EdgeBridgeGluingSpectrum_Theorem}, we introduce an algorithm that computes this spectrum.

\noindent\rule{15.24cm}{0.8pt} \\
\textbf{Algorithm 1}: Compute the spectra of the glued graph $\Gamma = \Gamma_1 \sqcup_B \Gamma_2$. \\
\noindent\rule{15.24cm}{0.8pt} \\
\textbf{Input}: Two disjoint graphs $\Gamma_1$ and $\Gamma_2$, and a set of $k$ bridges $\{e_1, \ldots, e_k\}$. \\
\begin{tabular}{l l l l}
& & & \textit{\underline{Note}: Bridge $e_i$ has endpoints $v^1_i \in V(\Gamma_1)$ and $v^2_i \in V(\Gamma_2)$}.
\vspace{2mm}
\end{tabular} \\
\textbf{Output}: The characteristic polynomial of the glued graph $\Gamma = \Gamma_1 \sqcup_B \Gamma_2$. \\
\underline{Step 1}: Connect $\Gamma_1$ and $\Gamma_2$ using edge $e_1$. Update the characteristic polynomial $p_{\Delta_{\Gamma}}(\lambda)$ and the glued graph $\Gamma$. \\
\begin{tabular}{l l}
& $p_{\Delta_{\Gamma}}(\lambda) = p_{\Delta_{\Gamma_1}}(\lambda) p_{\Delta_{\Gamma_2}}(\lambda) + p_{\Delta_{\Gamma_1}}(\lambda) p_{\Delta_{\Gamma_2 \left(v^2_1, v^2_1\right)}}(\lambda) + p_{\Delta_{\Gamma_1 \left(v^1_1, v^1_1\right)}}(\lambda) p_{\Delta_{\Gamma_2}}(\lambda)$ (\textbf{Theorem~\ref{EdgeBridgeGluingSpectrum_Theorem}}), \\
& $\Gamma = \{\Gamma_1 \sqcup_{B_1} \Gamma_2 | V(B_1) = \{v^1_1, v^2_1\}, E(B_1) = \{e_1\} \}$.
\vspace{2mm}
\end{tabular} \\
\underline{Step 2}: Iteratively add the edges $\{e_2, \ldots, e_k\}$, while updating $p(\lambda)$ and $\Gamma$. \\
\textbf{for each} $\{e_i \in \{e_2, \ldots, e_k\} \}$ \textbf{do} \\
\begin{tabular}{l | l}
& $p_{\Delta_{\Gamma}}(\lambda) = p_{\Delta_{\Gamma}}(\lambda) + p_{\Delta_{\Gamma \left(v^{1}_i, v^{1}_i\right)}}(\lambda) + p_{\Delta_{\Gamma \left(v^{2}_i, v^{2}_i\right)}}(\lambda)$ \\
& \quad \quad \quad \quad $ - 2 (-1)^{v^{1}_i+v^{2}_i} \det \big( \left( \Delta_{\Gamma} - \lambda I \right)_{(v^{1}_i, v^{2}_i)} \big)$
(\textbf{Theorem~\ref{EdgeInterfaceSpectrum_Theorem}}), \\
& $\Gamma = \{\Gamma \sqcup_{I_i} P_2 | V(I_i) = \{v^1_i, v^2_i\} = V(P_2), E(I_i) = \{e_i\} \} $. \\
\end{tabular} \\
\vspace{2mm}
\textbf{end} \\
\textbf{Return}: $p_{\Delta_{\Gamma}}(\lambda)$. \\
\noindent\rule{15.24cm}{0.8pt}

Note that this algorithm is parallelizable, i.e. we could assign to different processors each characteristic polynomial term in the summations. A faster computation of the eigenvalues of the Laplacian matrix is therefore possible, which is especially useful in cases of large graphs. This is a suggested topic for future work, as discussed in Section \ref{Conclusion_Section}.


\section{Examples} \label{Examples_Section}

\subsection{Complete Graph Gluing} \label{CompleteGraphGluingExample_Subsection}

We apply our results to the special case of gluing two complete graphs, which results in simple computations of the even and odd Laplacians, spectrum, and Fiedler value. This illustrates some useful descriptions in network theory, which can be easily derived from the relatively straightforward even and odd Laplacian matrices.

Note that the even Laplacian for an $n$-vertex complete graph $K_n$ is an $n \times n$-matrix given by:
\begin{equation*}
\Delta^{+}_{K_n}(i,j) = \begin{cases}
n-1 & \text{if } i=j \\
-1 & \text{otherwise,}
\end{cases}
\end{equation*}
and the odd Laplacian is a matrix of the size $\frac{n(n-1)}{2} \times \frac{n(n-1)}{2}$ given by:
\begin{equation*}
\Delta^{-}_{K_n}(i,j) = \begin{cases}
2 & \text{if } i=j \\
\zeta & \text{if } i \neq j,
\end{cases}
\end{equation*}
where 
\begin{equation*}
\zeta=\begin{cases}
0 & \text{if } e_i \text{ is not incident to } e_j\\
1 & \text{if } e_i \text{ is incident to } e_j \text{ at } v_s \text{, and } e_i \text{ and } e_j \text{ both start or end at } v_s\\
-1 & \text{if } e_i \text{ is incident to } e_j \text{ at } v_s \text{, and one of the edges starts at } v_s \text{, the other ends at } v_s.
\end{cases}
\end{equation*}

We begin by applying the interface and bridge gluing formulae for the Laplacian matrices of $K_n$.

\begin{corollary}\label{CompleteGraphInterfaceGluing_Corollary}
Let $K_m$ and $K_n$ be two complete graphs, and $\Gamma_{1+n-q}$ is the graph obtained by the interface gluing of $K_m$ and $K_n$. Suppose 
$$V(K_m)=\{v_1, \ldots, v_{m-q+1}, \ldots, v_m\},$$
and 
$$V(K_n)=\{v_{m-q+1}, \ldots, v_m, \ldots, v_{m-q+n}\},$$ and let $I$ be the interface such that $V(I)=\{v_{m-q+1},\ldots,v_m\}$. Then the even Laplacian of $\Gamma$ is an $(m+n-q) \times (m+n-q)$-matrix given by:
\begin{equation*}
\Delta^+_{\Gamma}(i,j) = \begin{cases}
m-1 & \text{if } i=j \leq m-q\\
n-1 & \text{if } i=j>m\\
m+n-2 & \text{if } m-q < i=j \leq m\\
-1 & \text{if } i,j \leq m \text{ or } i,j \geq m-q+1\\
0 & \text{otherwise,}
\end{cases}
\end{equation*}
and the odd Laplacian of $\Gamma$ is a matrix of the size $\frac{m(m-1)+n(n-1)-q(q-1)}{2} \times \frac{m(m-1)+n(n-1)-q(q-1)}{2}$ given by:
\begin{equation*}
\Delta^-_{\Gamma}(i,j) = \begin{cases}
2 & \text{if } i=j\\
\zeta & \text{if } i \neq j,
\end{cases}
\end{equation*}
where 
\begin{equation*}
\zeta=\begin{cases}
0 & \text{if } e_i \text{ is not incident to } e_j\\
1 & \text{if } e_i \text{ is incident to } e_j \text{ at } v_s \text{, and } e_i \text{ and } e_j \text{ both start or end at } v_s\\
-1 & \text{if } e_i \text{ is incident to } e_j \text{ at } v_s \text{, and one of the edges starts at } v_s \text{, the other ends at } v_s. 
\end{cases}
\end{equation*}
\end{corollary}

\begin{corollary}\label{CompleteGraphBridgeGluing_Corollary}
Let $K_m$ and $K_n$ be two complete graphs, and $\Gamma_{1+n}$ be the graph obtained by the bridge gluing of $K_m$ and $K_n$. Suppose $V(K_m)=\{v_1, \ldots, v_m\}$ and $V(K_n)=\{v_{m+1}, \ldots, v_{m+n}\}$, and let $B$ be the bridge graph. Then the even Laplacian of $\Gamma$ is an $(m+n) \times (m+n)$-matrix given by:
\begin{equation*}
\Delta^+_{\Gamma}(i,j) = \begin{cases}
m-1 & \text{if } i=j \leq m \text{ and } v_i \notin B\\
m & \text{if } i=j \leq m \text{ and } v_i \in B\\
n-1 & \text{if } i=j>m \text{ and } v_i \notin B\\
n & \text{if } i=j>m \text{ and } v_i \in B\\
-1 & \text{if } i,j \leq m \text{ or } i,j>m\\
0 & \text{otherwise,}
\end{cases}
\end{equation*}
and the odd Laplacian of $\Gamma$ is a matrix of the size $(\frac{m(m-1)+n(n-1)}{2}+|E(B)|) \times (\frac{m(m-1)+n(n-1)}{2}+|E(B)|)$ given by:
\begin{equation*}
\Delta^-_{\Gamma}(i,j) = \begin{cases}
2 & \text{if } i=j\\
\zeta & \text{if } i \neq j,
\end{cases}
\end{equation*}
where 
\begin{equation*}
\zeta=\begin{cases}
0 & \text{if } e_i \text{ is not incident to } e_j\\
1 & \text{if } e_i \text{ is incident to } e_j \text{ at } v_s \text{, and } e_i \text{ and } e_j \text{ both start or end at } v_s\\
-1 & \text{if } e_i \text{ is incident to } e_j \text{ at } v_s \text{, and one of the edges starts at } v_s \text{, the other ends at } v_s.
\end{cases}
\end{equation*}
\end{corollary}

The proofs of Corollary~\ref{CompleteGraphInterfaceGluing_Corollary} and Corollary~\ref{CompleteGraphBridgeGluing_Corollary} follow exactly the same as Theorem~\ref{EvenLaplacianGluing_Theorem} and Theorem~\ref{OddLaplacianGluing_Theorem}.

We also provide explicit consequences of our results to computing the spectrum of a graph obtained from gluing two complete graphs.

\begin{corollary} \label{CompleteGraphGluingSpectrum_Corollary}
Let $K_m$ and $K_n$ be two complete graphs with $m$ and $n$ vertices respectively. Choose two vertices $v_m \in V(K_m)$ and $v_n \in V(K_n)$ to define an interface $I$ where $V(I) = \{v_m, v_n\}$ and $E(I) = \emptyset$. Then, the characteristic polynomial of $\Delta_{K_m \sqcup_I K_n}$ is
\begin{equation*}
p_{\Delta_{K_m \sqcup_I K_n}}(\lambda) = (-1)^{m+n-1}\lambda(\lambda-m)^{m-2}(\lambda-n)^{n-2}(\lambda-1)(\lambda-(m+n-1)).
\end{equation*}
If instead we define a bridge graph $B$ where $V(B) = \{v_m, v_n\}$ and $E(B) = \{e\}$ where $e$ has endpoints $v_m$ and $v_n$, then the characteristic polynomial of $\Delta_{K_m \sqcup_B K_n}$ is
\begin{equation*}
p_{K_m \sqcup_B K_n}(\lambda) = (-1)^{m+n}\lambda(\lambda-m)^{m-2}(\lambda-n)^{n-2}q_{m,n}(\lambda),
\end{equation*}
where
\begin{equation*}
q_{m,n}(\lambda)= \lambda^3 -(m+n+2)\lambda^2+(1+(m+1)(n+1))\lambda - (m+n).
\end{equation*}
\end{corollary}

The derivation for both characteristic polynomials can be done using Lemmas~\ref{CompleteGraphPolynomial_Lemma} and~\ref{CompleteGraphPolynomialTruncated_Lemma}, and $p_{\Delta_{K_m \sqcup_I K_n}}(\lambda)$ follows from Theorem~\ref{VertexInterfaceSpectrum_Theorem} while $p_{\Delta_{K_m \sqcup_B K_n}}(\lambda)$ follows from Theorem~\ref{EdgeBridgeGluingSpectrum_Theorem}. As a consequence of Corollary~\ref{CompleteGraphGluingSpectrum_Corollary}, we can also determine the number of spanning trees of a graph obtained by gluing two complete graphs either via a single vertex or with a single edge.

\begin{proposition} \label{SpanningTreeCount_Proposition}
Suppose we glue two complete graphs $K_m$ and $K_n$ via an interface such that the interface $I$ is defined by $V(I) = \{v_m\} = \{v_n\}$ where $v_m \in V(K_m)$ and $v_n \in V(K_n)$. The number of spanning trees in $K_m \sqcup_I K_n$ then is $m^{m-2} n^{n-2}$.

Furthermore, suppose we glue $K_m$ and $K_n$ via a bridge graph such that the bridge graph $B$ is defined by $V(B) = \{v_m, v_n\}$ where $v_m \in V(K_m)$ and $v_n \in V(K_n)$, and $E(B) = \{e\}$ where $e$ is an edge with endpoints $v_m$ and $v_n$. Then, $K_m \sqcup_B K_n$ has exactly $m^{m-2} n^{n-2}$ spanning trees as well.
\end{proposition}

\begin{proof}
By Kirchoff's Theorem, the number of spanning trees $T(\Gamma)$ for a connected graph $\Gamma$ with $n$ vertices is given by
\begin{equation*}
T(\Gamma)= \frac 1 n \prod_{i=2}^n \lambda_i,
\end{equation*}
where $0 < \lambda_2 \leq \ldots \leq \lambda_n$ are the ordered eigenvalues of $\Delta_{\Gamma}$. By Corollary~\ref{CompleteGraphGluingSpectrum_Corollary}, the nonzero eigenvalues of $\Delta_{K_m \sqcup_I K_n}$ are $m$ (with multiplicity $m-2$), $n$ (with multiplicity $n-2$), $1$, and $m+n-1$. Therefore,
\begin{equation*}
T(K_m \sqcup_I K_n) = \frac{1}{m+n-1} m^{m-2} n^{n-2} (m+n-1) = m^{m-2} n^{n-2},
\end{equation*}
as we wanted. Similarly, since the nonzero eigenvalues of $\Delta_{K_m \sqcup_B K_n}$ are $m$ (with multiplicity $m-2$), $n$ (with multiplicity $n-2$), and the zeros of $q_{m,n}(\lambda)$, whose product is the constant term of $q_{m,n}(\lambda)$, it follows that
\begin{equation*}
T(K_m \sqcup_B K_n)= \frac{1}{|V(K_m \sqcup_B K_n)|} \prod_{i=2}^{|V(K_m \sqcup_B K_n)|} \lambda_i = \frac{1}{m+n} m^{m-2} n^{n-2} (m+n)=m^{m-2}n^{n-2},
\end{equation*}
as we had expected.
\end{proof}

\begin{remark} Another way to derive Proposition~\ref{SpanningTreeCount_Proposition} is to realize that 1) the number of spanning trees in $K_n$ is $n^{n-2}$, and 2) the number of spanning trees is multiplicative with respect to the gluing, since a tree in both $K_m \sqcup_I K_n$ and $K_m \sqcup_B K_n$ is uniquely determined by choosing a tree in $K_m$ and a tree in $K_n$.
\end{remark}

We also provide some implications of Corollary~\ref{CompleteGraphGluingSpectrum_Corollary} on the Fiedler value of graphs (i.e. the smallest positive eigenvalue of the graph Laplacian) obtained by gluing complete graphs.

\begin{corollary} \label{CompleteInterfaceFiedler_Corollary}
Let $m,n \geq 2$. Suppose we have two complete graphs $K_m$ and $K_n$, and choose $v_m \in K_m$ and $v_n \in K_n$. If we glue $K_m$ and $K_n$ via an interface $I$, where $V(I) = \{v_m\} = \{v_n\}$, then the Fiedler value of $K_m \sqcup_I K_n$ is $1$.
\end{corollary}

\begin{proof}
By Corollary~\ref{CompleteGraphGluingSpectrum_Corollary}, the lowest nonzero eigenvalue necessarily has to be $1$.
\end{proof}

\begin{corollary} \label{CompleteBridgeDiffVerticesFiedler_Corollary}
Let $m,n \geq 2$. Suppose we have two complete graphs $K_m$ and $K_n$, and choose $v_m \in K_m$ and $v_n \in K_n$. If we glue $K_m$ and $K_n$ via a bridge graph $B$ such that $V(B) = \{v_m, v_n\}$ and $E(B) = \{e\}$ where the edge $e$ has endpoints $v_m$ and $v_n$, then the Fiedler value of $K_m \sqcup_B K_n$ satisfies the inequalities
\begin{equation*}
\min \{m,n, \frac 1 3 (m+n+2-2\sqrt{m^2+n^2-mn+m+n-6}) \}\leq F,
\end{equation*}
and
\begin{equation*}
F \leq \min \{m,n, \frac 1 3 (m+n+2+2\sqrt{m^2+n^2-mn+m+n-6}) \}.
\end{equation*}
\end{corollary}

\begin{proof}
From Corollary~\ref{CompleteGraphGluingSpectrum_Corollary}, since the zeroes of $q_{m,n}(\lambda)$ are real, we can apply the bounds of the zeroes of Laguerre polynomials \cite{1880_Laguerre}, so that the lowest nonzero eigenvalues lies within these bounds and the minimum between $m$ and $n$.
\end{proof}

We can sharpen these bounds to an equality when $m=n$.

\begin{corollary} \label{CompleteBridgeSameVerticesFiedler_Corollary}
Suppose $n \geq 2$ and we glue two $n$-vertex complete graphs $K^1_n$ and $K^2_n$ via a bridge graph $B$ such that $V(B) = \{v^1, v^2\}$ and $E(B) = \{e\}$ where $e$ has endpoints $v^1$ and $v^2$. Then, the Fiedler value of $K^1_n \sqcup_B K^2_n$ is
\begin{equation*}
F= \frac{1}{2} (n+2-\sqrt{n^2+4n-4}).
\end{equation*}
\end{corollary}

\begin{proof}
Similarly, the characteristic polynomial of $\Delta_{K^1_n \sqcup_B K^2_n}$ is
\begin{equation*}
p_{\Delta_{K^1_n \sqcup_B K^2_n}}(\lambda) = \lambda (\lambda-n)^{2n - 4} q_{n,n}(\lambda),
\end{equation*}
where
\begin{equation*}
\begin{split}
q_{n,n}(\lambda) & = \lambda^3-(2n+2)\lambda^2+(1+(n+1)^2)\lambda -2n \\
& = (\lambda - 1) (\lambda^2 - (n+2) \lambda + 2).
\end{split}
\end{equation*}
The zeroes of $p_{\Delta_{K^1_n \sqcup_B K^2_n}}(\lambda)$ are therefore $0$, $1$, $n$, and $\frac{1}2 (n+2 \pm \sqrt{n^2+4n-4})$. Since \begin{equation*}
0 < \frac{1}{2} (n+2-\sqrt{n^2+4n-4})< 1,
\end{equation*}
for all $n > 1$, this concludes the proof.
\end{proof}


\subsection{Path Graph Gluing} \label{PathGraphGluingExample_Subsection}

We provide another application of our results to the context of gluing path graphs. This holds potential in describing graph quantum mechanics (Section~\ref{QuantumMechanics_Subsubsection}), where we can extend the 1D model of a wave function existing on $\mathbb{R}$ to a 2D model existing on $\mathbb{R}^2$.

A \emph{path graph} or \emph{linear graph} is a graph whose vertices can be listed in the order $v_1, \hdots, v_n$ such that the edges are $\{v_i, v_{i+1}\}$ where $i = 1, \hdots, n-1$.
The even Laplacian for any $n$-vertex path graph $P_n$ is always an $n \times n$-matrix given by:
\begin{equation*}
\Delta^+_{P_n}(i,j) = \begin{cases}
1 & \text{if } i=j=1 \text{ or } i=j=n\\
2 & \text{if } 1 < i=j < n\\
-1 & \text{if } |i-j|=1\\
0 & \text{if } |i-j|>1,
\end{cases}
\end{equation*}
and the odd Laplacian is always an $(n-1) \times (n-1)$-matrix given by:
\begin{equation*}
\Delta^-_{P_n}(i,j) = \begin{cases}
2 & \text{if } i=j\\
0 & \text{if } |i-j|>1\\
1 & \text{if } |i-j| = 1 \text{ and } e_i, e_j \text{ both start or end at the common vertex} \\
-1 & \text{if } |i-j| = 1 \text{ and the common vertex is the start point of one and the end point of the other.}
\end{cases}
\end{equation*}

Notice that the interface gluing or bridge gluing of two path graphs to obtain a new path graph changes only the size of the even and odd Laplacians.

\begin{corollary}\label{PathGraphInterfaceGluing_Corollary}
Let $P_m$ and $P_{n-m+q}$ be two path graphs, and $P_n$ is the path graph obtained by the interface gluing of $P_m$ and $P_{n-m+q}$. Suppose $V(P_m)=\{v_1, \ldots, v_{m-q+1}, \ldots, v_m\}$ and $V(P_{n-m+q})=\{v_{m-q+1}, \ldots, v_m, \ldots, v_n\}$, and let $I$ be the interface such that $V(I)=\{v_{m-q+1},\ldots,v_m\}$. Then the even Laplacian of $P_n$ is an $n \times n$-matrix given by:
\begin{equation*}
\Delta^+_{P_n}(i,j) = \begin{cases}
1 & \text{if } i=j=1 \text{ or } i=j=n\\
2 & \text{if } 1 < i=j < n\\
-1 & \text{if } |i-j|=1\\
0 & \text{if } |i-j|>1,
\end{cases}
\end{equation*}
and the odd Laplacian of $P_n$ is an $(n-1) \times (n-1)$-matrix given by:
\begin{equation*}
\Delta^-_{P_n}(i,j) = \begin{cases}
2 & \text{if } i=j\\
0 & \text{if } |i-j|>1\\
1 & \text{if } |i-j| = 1 \text{ and } e_i, e_j \text{ both start or end at the common vertex} \\
-1 & \text{if } |i-j| = 1 \text{ and the common vertex is the start point of one and the end point of the other.}
\end{cases}
\end{equation*}
\end{corollary}

\begin{corollary}\label{PathGraphBridgeGluing_Corollary}
Let $P_m$ and $P_{n-m-q-1}$ be two path graphs, and let $P_n$ be the path graph obtained by the bridge gluing of $P_m$ and $P_{n-m-q-1}$. Suppose 
$$V(P_m)=\{v_1, \ldots, v_m\},$$
and
$$V(P_{n-m-q-1})=\{v_{m+q}, v_{m+q+1}, \ldots, v_n\},$$ 
and let the bridge graph $B$ be a path graph such that $V(B)=\{v_m,\ldots,v_{m+q}\}$. Then the even Laplacian of $P_n$ is an $n \times n$-matrix given by:
\begin{equation*}
\Delta^+_{P_n}(i,j) = \begin{cases}
1 & \text{if } i=j=1 \text{ or } i=j=n\\
2 & \text{if } 1 < i=j < n-q\\
-1 & \text{if } |i-j|=1\\
0 & \text{if } |i-j|>1,
\end{cases}
\end{equation*}
and the odd Laplacian of $P_n$ is an $(n-1) \times (n-1)$-matrix given by:
\begin{equation*}
\Delta^-_{P_n}(i,j) = \begin{cases}
2 & \text{if } i=j\\
0 & \text{if } |i-j|>1\\
1 & \text{if } |i-j| = 1 \text{ and } e_i, e_j \text{ both start or end at the common vertex} \\
-1 & \text{if } |i-j| = 1 \text{ and the common vertex is the start point of one and the end point of the other.}
\end{cases}
\end{equation*}
\end{corollary}

The proofs of Corollary~\ref{PathGraphInterfaceGluing_Corollary} and Corollary~\ref{PathGraphBridgeGluing_Corollary} follow exactly the same as Theorem~\ref{EvenLaplacianGluing_Theorem} and Theorem~\ref{OddLaplacianGluing_Theorem}.


\subsection{Cycle Graph Gluing} \label{CycleGraphGluingExample_Subsection}
A \emph{cycle graph} is a path graph with the added edge $\{v_n, v_1\}$. The even Laplacian for any $n$-vertex cycle graph $C_n$ is an $n \times n$-matrix given by:
\begin{equation*}
\Delta^{+}_{C_n}(i,j) = \begin{cases}
2 & \text{if } i=j \\
-1 & \text{if } |i-j|=1 \text{ or } (i,j) \in \{(1,n),(n,1)\} \\
0 & \text{otherwise,}
\end{cases}
\end{equation*}
and the odd Laplacian is an $n \times n$-matrix given by:
\begin{equation*}
\Delta^{-}_{C_n}(i,j) = \begin{cases}
2 & \text{if } i=j\\
0 & \text{if } 1 < |i-j| < n-1\\
1 & \text{if } |i-j| = 1 \text{ or } (i,j) \in \{(1,n-1),(n-1,1)\} \text{ and } e_i, e_j \text{ both start or end at the common vertex} \\
-1 & \text{otherwise.}
\end{cases}
\end{equation*}

\begin{corollary}\label{CycleGraphInterfaceGluing_Corollary}
Let $C_m$ and $C_{n-m+q}$ be two cycle graphs, and $\Gamma$ is the graph obtained by the interface gluing of $C_m$ and $C_{n-m+q}$. Suppose $V(C_m)=\{v_1, \ldots, v_{m-q+1}, \ldots, v_m\}$ and $V(C_{n-m+q})=\{v_{m-q+1}, \ldots, v_m, \ldots, v_n\}$, and let $I$ be the interface such that $V(I)=\{v_{m-q+1},\ldots,v_m\}$. Then the even Laplacian of $\Gamma$ is an $n \times n$-matrix given by:
\begin{equation*}
\Delta^+_{\Gamma}(i,j) = \begin{cases}
3 & \text{if } i=j=m-q+1 \text{ or } i=j=m\\
2 & \text{if } i=j \text{ and } i,j \neq m-q+1 \text{ and } i,j \neq m\\
-1 & \text{if } |i-j|=1 \text{ or } (i,j) \in \{(1,m),(m,1),(m-q+1,n),(n,m-q+1)\} \\
0 & \text{otherwise,}
\end{cases}
\end{equation*}
and the odd Laplacian of $\Gamma$ is an $(n-2) \times (n-2)$-matrix given by:
\begin{equation*}
\Delta^-_{\Gamma}(i,j) = \begin{cases}
2 & \text{if } i=j\\
0 & \text{if } |i-j| > 1 \text{ and } (i,j) \notin \mathcal{S} \\
1 & \text{if } |i-j| = 1 \text{ or } (i,j) \in \mathcal{S} \text{ and } e_i, e_j \text{ both start or end at the common vertex} \\
-1 & \text{otherwise,}
\end{cases}
\end{equation*}
where $\mathcal{S}=\{(1,m-1),(m-1,1),(n-m+q,n-1),(n-1,n-m+q)\}$.
\end{corollary}

\begin{corollary}\label{CycleGraphBridgeGluing_Corollary}
Let $C_m$ and $C_n$ be two cycle graphs, and $\Gamma$ is the graph obtained by the bridge gluing of $C_m$ and $C_n$. Suppose $V(C_m)=\{v_1, \ldots, v_m\}$ and $V(C_n)=\{v_{m+1}, v_{m+2}, \ldots, v_{m+n}\}$, and let $B$ be the bridge graph. Then the even Laplacian of $\Gamma$ is an $(m+n) \times (m+n)$-matrix given by:
\begin{equation*}
\Delta^+_{\Gamma}(i,j) = \begin{cases}
3 & \text{if } i=j \text{ and } v_i \in B\\
2 & \text{if } i=j \text{ and } v_i \notin B\\
-1 & \text{if } v_i \text{ is adjacent to } v_j\\
0 & \text{otherwise,}
\end{cases}
\end{equation*}
and the odd Laplacian of $\Gamma$ is a matrix of the size $(m+n-2+|E(B)|) \times (m+n-2+|E(B)|)$ given by:
\begin{equation*}
\Delta^-_{\Gamma}(i,j) = \begin{cases}
2 & \text{if } i=j\\
\zeta & \text{if } i \neq j,
\end{cases}
\end{equation*}
where 
\begin{equation*}
\zeta=\begin{cases}
0 & \text{if } e_i \text{ is not incident to } e_j\\
1 & \text{if } e_i \text{ is incident to } e_j \text{ at } v_s \text{, and } e_i \text{ and } e_j \text{ both start or end at } v_s\\
-1 & \text{if } e_i \text{ is incident to } e_j \text{ at } v_s \text{, and one of the edges starts at } v_s \text{, the other ends at } v_s.
\end{cases}
\end{equation*}
\end{corollary}

The proofs of Corollary~\ref{CycleGraphInterfaceGluing_Corollary} and Corollary~\ref{CycleGraphBridgeGluing_Corollary} follow exactly the same as Theorem~\ref{EvenLaplacianGluing_Theorem} and Theorem~\ref{OddLaplacianGluing_Theorem}.

We can also explicitly compute the spectrum of a graph obtained by gluing two cycle graphs.

\begin{corollary}
Let $C_m$ and $C_n$ be two cycle graphs with $m$ and $n$ vertices respectively. Choose two vertices $v_m \in V(C_m)$ and $v_n \in V(C_n)$ to define an interface $I$ where $V(I) = \{v_m, v_n\}$. Then, the characteristic polynomial of $\Delta_{C_m \sqcup_I C_n}$ is
\begin{equation*}
\begin{split}
p_{\Delta_{C_m \sqcup_I C_n}}(\lambda) & = \prod_{i=0}^{m-1} \prod_{j=1}^{n-1} \Bigg( 4\left[ 1 - \cos\left(\frac{2 \pi i}{m}\right) \right] \left[ 1 - \cos\left(\frac{\pi j}{n}\right)\right] + 2 \lambda \left[ \cos\left(\frac{2 \pi i}{m}\right) + \cos\left(\frac{\pi j}{n}\right) - 2\right] + \lambda^2 \Bigg) \\
& + \prod_{i=1}^{m-1} \prod_{j=0}^{n-1} \Bigg( 4\left[ 1 - \cos\left(\frac{\pi i}{m}\right) \right] \left[ 1 - \cos\left(\frac{2 \pi j}{n}\right)\right] + 2 \lambda \left[ \cos\left(\frac{\pi i}{m}\right) + \cos\left(\frac{2 \pi j}{n}\right) - 2\right] + \lambda^2 \Bigg) \\
& + \lambda \prod_{i=1}^{m-1} \prod_{j=1}^{n-1} \Bigg( 4\left[ 1 - \cos\left(\frac{\pi i}{m}\right) \right] \left[ 1 - \cos\left(\frac{\pi j}{n}\right)\right] + 2 \lambda \left[ \cos\left(\frac{\pi i}{m}\right) + \cos\left(\frac{\pi j}{n}\right) - 2\right] + \lambda^2 \Bigg).
\end{split}
\end{equation*}
Furthermore, if we define a bridge graph $B$ where $V(B) = \{v_m, v_n\}$ and $E(B) = \{e\}$ where $e$ has endpoints $v_m$ and $v_n$, then the characteristic polynomial of $\Delta_{C_m \sqcup_B C_n}$ is
\begin{equation*}
\begin{split}
p_{\Delta_{C_m \sqcup_B C_n}}(\lambda) & = 
\prod_{i=0}^{m-1} \prod_{j=0}^{n-1} \Bigg( 4\left[ 1 - \cos\left(\frac{2 \pi i}{m}\right) \right] \left[ 1 - \cos\left(\frac{2 \pi j}{n}\right) \right] + 2 \lambda \left[ \cos\left(\frac{2 \pi i}{m}\right) + \cos\left(\frac{2 \pi j}{n}\right) - 2\right] + \lambda^2 \Bigg) \\
& + \prod_{i=0}^{m-1} \prod_{j=1}^{n-1} \Bigg( 4\left[ 1 - \cos\left(\frac{2 \pi i}{m}\right) \right] \left[ 1 - \cos\left(\frac{\pi j}{n}\right)\right] + 2 \lambda \left[ \cos\left(\frac{2 \pi i}{m}\right) + \cos\left(\frac{\pi j}{n}\right) - 2\right] + \lambda^2 \Bigg) \\
& + \prod_{i=1}^{m-1} \prod_{j=0}^{n-1} \Bigg( 4\left[ 1 - \cos\left(\frac{\pi i}{m}\right) \right] \left[ 1 - \cos\left(\frac{2 \pi j}{n}\right)\right] + 2 \lambda \left[ \cos\left(\frac{\pi i}{m}\right) + \cos\left(\frac{2 \pi j}{n}\right) - 2\right] + \lambda^2 \Bigg).
\end{split}
\end{equation*}
\end{corollary}
The derivation for these characteristic polynomials follow a similar form as Corollary~\ref{CompleteGraphGluingSpectrum_Corollary}, only instead we use Lemmas~\ref{CycleGraphPolynomial_Lemma} and~\ref{CycleGraphPolynomialTruncated_Lemma}.


\section{Perspectives and Future Work} \label{Conclusion_Section}
We provide formulae to generate the even and odd Laplacians of the interface and bridge gluing, as well as the spectrum of the Laplacian of the glued graph as a function of the spectra of the original graphs. We also provide an algorithm (Section~\ref{Algorithm_Subsection}) that uses these theorems to compute the spectrum of a graph obtained by gluing two graph via a bridge graph $B$ where $|E(B)| > 1$. This suggests that improvements in computational efficiency when studying the spectra of large graphs are feasible with this approach.

We outline some directions for future work, particularly in network theory, graph quantum mechanics, and parallelizable electronic structure calculations.

\subsection{Network Theory}
Motivated by applications to network theory, one prospective direction of this work involves developing a gluing formula for the spectrum of the normalized Laplacian matrix $\Delta_{\Gamma}^{norm}$, from which we may derive sharper bounds for the Cheeger constant. Additionally, we plan to investigate the relationship between the Fiedler value and the Cheeger constant using the definition of the Laplacian matrix used in this manuscript.

The Cheeger constant \cite{1969_Cheeger, 1987_Mohar} is a quantity that is relevant while studying bottlenecks on graphs. The quantity is defined by the following:

\begin{definition} \label{CheegerConstant_Definition}
The Cheeger constant $h(\Gamma)$ of a connected graph $\Gamma$ is
\begin{equation*}
h(\Gamma) = \underset{X}{\min} \bigg\{ \frac{|\partial X|}{|X|} : X \subseteq V(\Gamma), 0 < |X| < \frac{1}{2} |V(\Gamma)| \bigg\}.
\end{equation*}
\end{definition}

A significant amount of research in network theory is dedicated towards understanding the relationship between the Cheeger constant and the Fiedler value, the relationship formally known as Cheeger inequalities \cite{2006_Hoory}. 
\begin{theorem} \label{CheegerInequality_Theorem}
Given a connected graph $\Gamma$ such that $F$ is the Fiedler value from the normalized Laplacian matrix and $h(\Gamma)$ is the Cheeger constant, it follows that
\begin{equation*}
2 h(\Gamma) \geq F \geq \frac{h(\Gamma)^2}{2}.
\end{equation*}
\end{theorem}

Our results on the particular classes of graphs suggest Cheeger inequality-like bounds for the number of bottlenecks in a graph. Prospective work will involve developing a gluing formalism for the normalized Laplacian matrix, from which we will provide some insights towards bottleneck detection in general graphs. 

\subsection{Graph Quantum Mechanics} \label{QuantumMechanics_Subsubsection}

The Schr\"{o}dinger equation
\begin{equation*}\label{Schrodinger_Equation}
i \hbar \frac{\partial}{\partial t} \Psi(t, x) = \bigg(-\frac{\hbar^2}{2m} \nabla^2 + \hat{V}\bigg) \Psi(t, x),
\end{equation*}
is one of the pillars of quantum mechanics that describes the evolution of quantum states. The Feynman path integral interpretation of quantum mechanics \cite{1965_Feynman} provides the time evolution of the wave function from time $t_0$ to $t$ in terms of the Green's function $K(t-t_0;x,y)$:
\begin{equation*}
\Psi(t, x) = \int_{y \in \mathbb{R}^n} K(t-t_0;x,y) \Psi(t_0, y) d^n y.
\end{equation*}
By using Witten's approach to supersymmetry on quantum mechanics \cite{1982_Witten}, we can relate the spectrum of the Laplacian to the topology of the configuration space.
There have been several efforts in developing a graph-theoretic analogue of quantum mechanics, see e.g. \cite{2012_DelVecchio, 2007_Mnev, 2016_Mnev, 2016_Wong}. It relies on the fact that the Laplacian matrix is a discretized version of the Laplace operator
\begin{equation*}
\Delta = \nabla^2 = \sum_{i=1}^{n} \frac{\partial^2}{\partial x_i^2},
\end{equation*}
and a particular instance of the Hodge-Laplace operator
\begin{equation*}
\Delta = d d^* + d^* d,
\end{equation*}
with Euclidean metric on $\mathbb{R}^n$.

The partition function for the free particle can be used to count special types of walks on graphs \cite{2017_Yu_Hyperwalk, 2017_Yu_Superwalk} and to compute topological invariants \cite{2017_Xu}.
We can replace the continuous Laplacian operator $\nabla^2$ with the matrix $\Delta_{\Gamma}$ in situations obeying the principle of locality, such as a particle confined to an optical lattice \cite{2005_Bloch} where $\mathbb{R}^n$ can be approximated by a graph $\Gamma$. Following the formalism in \cite{2016_Mnev}, we study the discrete analogue of the free particle Schr\"{o}dinger equation

\begin{equation*}
i \frac{\partial}{\partial t} \Psi(t) = - \frac{\hbar}{2m} \Delta_{\Gamma} \Psi(t),
\end{equation*}
where $\Psi(t) \in \mathbb{C}^{|V|}$ is the wave function and $\Delta_{\Gamma}$ is the Laplacian matrix. The solution is
\begin{equation*}
\Psi(t) = e^{\frac{i \hbar}{2m} \Delta_{\Gamma} (t-t_0)} \Psi(t_0),
\end{equation*}
where $e^{\frac{i \hbar}{2m} \Delta_{\Gamma} (t-t_0)}$ is an operator whose entries are analogous to the propagator $K(t-t_0; x,y)$. In other words, the discretized description of the evolution of states on a graph $\Gamma$ can be expressed as
\begin{equation*}
\Psi(t, x) = \sum_{y \in V(\Gamma)} \braket{x | e^{\frac{i \hbar}{2m} \Delta_{\Gamma} (t-t_0)} | y} \Psi(t_0, y).
\end{equation*}
An explicit derivation for how the propagator $\braket{x | e^{\frac{i \hbar}{2m} \Delta_{\Gamma} (t-t_0)} | y}$ relates to the paths $\gamma$ from $y \in V(\Gamma)$ to $x \in V(\Gamma)$, where $\Gamma$ is a $k$-regular graph, is given in \cite{2016_Mnev}:
\begin{equation*}
\braket{x | e^{\frac{i \hbar}{2m} \Delta_{\Gamma} (t-t_0)} | y} = \sum_{\text{Paths } \gamma \text{ from } y \text{ to } x} \frac{(t-t_0)^{|\gamma|}}{|\gamma|!} e^{ \frac{i\hbar}{2m} (t-t_0)}.
\end{equation*}
This provides a discretized version of the partition function for quantum mechanics and a combinatorial interpretation of counting paths in the Feynman path integral formulation. However, given that an eigenbasis exists for $\Delta_{\Gamma}$, the propagator can also be written as
\begin{equation*}
\braket{x | e^{\frac{i \hbar}{2m} \Delta_{\Gamma} (t-t_0)} | y} = \braket{x | E e^{\frac{i \hbar}{2m} \Lambda (t-t_0)} E^{-1} | y},
\end{equation*}
where
\begin{equation*}
\Lambda(i,j) = \begin{cases}
\lambda_i & \text{if } i=j \\
0 & \text{otherwise,}
\end{cases}
\end{equation*}
and
\begin{equation*}
E = [v_1(\lambda_1), v_2(\lambda_2), \hdots, v_n(\lambda_n)],
\end{equation*}
where $v_i(\lambda_i)$ is the $i^{th}$ eigenvector of $\Delta_{\Gamma}$ corresponding to the $\lambda_i$ eigenvalue.

Now suppose a particle initially lives in the $m$-vertex graph $\Gamma_1$, and the space of states $\mathbb{C}^{m}$ expands by gluing $\Gamma_1$ and $\Gamma_2$ such that $\Psi(t) \in \mathbb{C}^{M}$, where $M = m + |V(\Gamma_2)|$. The time evolution of $\Psi(t)$ can then be quickly computed given some information about the eigenvectors of $\Delta_{\Gamma}$, requiring a better understanding of the gluing procedure for eigenvectors. Not only do we expect those results to enhance time evolution computations on large graphs, but we also expect them to reveal direct implications on counting the number of paths between vertices.

\subsection{Electronic Structure Calculations and Graph Quantum Mechanics}

Following the concept of graph quantum mechanics, we propose a graph-theoretic perspective of electronic structure that admits parallelizable computations. Based on the time-independent Schr\"{o}dinger equation for the free particle
\begin{equation*}
\Delta_{\Gamma} \Psi = E \Psi,
\end{equation*}
where $E$ is the total energy of the system and $\Psi$ is the wave function, the allowed energy levels are precisely the eigenvalues of $\Delta_{\Gamma}$. Revisiting the example where the space of states $\mathbb{C}^m$ on an $m$-vertex graph $\Gamma_1$ expands by gluing $\Gamma_1$ and $\Gamma_2$ via a bridge graph $B$, we have an iterative method to calculate the energy levels of the glued graph in Section \ref{Algorithm_Subsection}. Moreover, this calculation is parallelizable with respect to the fact that the term-by-term summation in Theorems \ref{EdgeInterfaceSpectrum_Theorem} and \ref{EdgeBridgeGluingSpectrum_Theorem} can be split to individual processors. We therefore suggest that our results highlight an efficient computing method for allowed energy levels of a free particle.

Future work in this field will involve comparing the computing speeds of multi-core calculations, as well as investigating cases outside of the free particle ($\hat{V} \neq 0$) This implication further necessitates a method for computing the eigenvectors of the glued graph.


\newpage
\begin{appendix}

\section{Some Useful Lemmas} \label{Lemma_Section}
The following are standard results in linear algebra that we use to prove our results in this manuscript.

\begin{lemma}\label{Eigenspace_Lemma}
If $A$ is a (real or complex-valued) matrix, then $AA^*$ and $A^*A$ have the same eigenvalues. Moreover, if $\lambda \neq 0$ is an eigenvalue for $AA^*$ with multiplicity $m$, then $\lambda$ is also an eigenvalue for $A^*A$ with multiplicity $m$.
\end{lemma}

\begin{proof}
Let $S^{AA^*}$ be the eigenspace of $AA^*$, and $S^{A^*A}$ be the eigenspace of $A^*A$ with eigenvalue $\lambda>0$. We want to show $S^{AA^*}$ and $S^{A^*A}$ are isomorphic. Let $v$ be an eigenvector of $AA^*$, we have $AA^*v=\lambda v$ where $\lambda$ is the corresponding eigenvalue. Then $A^*AA^*v=\lambda A^*v$, so that $A^*v$ is an eigenvector of $A^*A$. Similarly, if $w$ is an eigenvector of $A^*A$, then $Aw$ is an eigenvector of $AA^*$. Therefore, $A$ is a linear transformation $\phi: S^{A^*A} \rightarrow S^{AA^*}$, and $A^*$ is a linear transformation $\psi: S^{AA^*} \rightarrow S^{A^*A}$. It follows that $\phi \circ \psi = \psi \circ \phi = \lambda I$, where $\lambda$ is the corresponding eigenvalue. Hence $\phi$ and $\psi$ are invertible. Thus, $\text{dim}(S^{A^*A}) = \text{dim}(S^{AA^*})$. Therefore, $S^{AA^*}$ and $S^{A^*A}$ are isomorphic.
\end{proof}

\begin{lemma} \label{BlockTriangular_Lemma}
Let $M=\begin{bmatrix}A&C\\0&B \end{bmatrix}$ be a block triangular matrices, i.e. $A$ and $B$ are square matrices. Then
\begin{equation*}
\det M = \det A \det B.
\end{equation*}
\end{lemma}

\begin{lemma} \label{SplitColumn_Lemma}
Let $A= [v_1, v_2, \ldots, v_k, \ldots, v_n]$ be an $n\times n$-matrix with column vectors $v_i$ such that $v_k= u+w$, for some $k, u$ and $w$. Then
\begin{equation*}
det(A)= \det ([v_1, v_2, \ldots, u, \ldots, v_n])+\det ([v_1, v_2, \ldots, w, \ldots, v_n]).
\end{equation*}
\end{lemma}

The following result describes explicitly the spectrum of a circulant matrix \cite{2006_Gray}.

\begin{lemma} \label{Circulant_Lemma}
Let $C$ be a $n \times n$ circulant matrix, with circulant row vector $(c_0,c_1, \ldots, c_{n-1})$. Let $\omega$ be a primitive $n$-th rooth of unity. Then the eigenvalues of $C$ are given by
\begin{equation*}
\lambda_i = c_0 +c_1 \omega_i+c_2 \omega_i^2+\ldots + c_{n-1}\omega_i^{n-1},
\end{equation*}
where $\omega_i = \omega^i$.
\end{lemma}

Consider the even Laplacian matrix of the complete graph $\Delta_{K_n}$. From Lemma~\ref{Circulant_Lemma}, we can deduce the characteristic polynomials of both $\Delta_{K_n}$ and $\left(\Delta_{K_n}\right)_{(v, v)}$, as well as the characteristic polynomial of $\Delta_{C_n}$.

\begin{lemma} \label{CompleteGraphPolynomial_Lemma}
The characteristic polynomial of the Laplacian matrix of the complete graph $K_n$ is 
\begin{equation*}
p_{\Delta_{K_n}}(\lambda)=(-1)^n \lambda (\lambda-n)^{n-1}.
\end{equation*}
\end{lemma}

\begin{lemma} \label{CompleteGraphPolynomialTruncated_Lemma}
The characteristic polynomial of $\Delta_{K_n (v, v)}$, where $v \in V(K_n)$, is
\begin{equation*}
p_{\Delta_{K_n (v, v)}}(\lambda) = (-1)^{n-1} (\lambda-1) (\lambda-n)^{n-2}.
\end{equation*}
\end{lemma}

\begin{lemma} \label{CycleGraphPolynomial_Lemma}
The characteristic polynomial of the Laplacian matrix of the cycle graph $C_n$ is
\begin{equation*}
p_{\Delta_{C_n}}(\lambda) = \prod_{j=0}^{n-1} \left(2 \left[1 - \cos\left( \frac{2 \pi j}{n} \right)\right] - \lambda \right).
\end{equation*}
\end{lemma}

\begin{lemma} \label{CycleGraphPolynomialTruncated_Lemma}
The characteristic polynomial of $\Delta_{C_n (v, v)}$, where $v \in V(C_n)$, is
\begin{equation*}
p_{\Delta_{C_n (v, v)}}(\lambda) = \prod_{j=1}^{n-1} \left(2 \left[1 - \cos\left(\frac{\pi j}{n}\right) \right] - \lambda\right).
\end{equation*}
\end{lemma}

\end{appendix}


\newpage
IVAN CONTRERAS\\
DEPARTMENT OF MATHEMATICS\\
UNIVERSITY OF ILLIONIS URBANA-CHAMPAIGN, IL\\
\textit{E-mail address: } \texttt{icontrer@illinois.edu}
\newline

MICHAEL TORIYAMA\\
DEPARTMENT OF MATHEMATICS, DEPARTMENT OF MATERIALS SCIENCE AND ENGINEERING\\
UNIVERSITY OF ILLIONIS URBANA-CHAMPAIGN, IL\\
\textit{E-mail address: }\texttt{toriyam2@illinois.edu}
\newline

CHENGZHENG YU\\
DEPARTMENT OF ECONOMICS\\
GEORGETOWN UNIVERSITY, D.C.\\
\textit{E-mail address: }\texttt{cy375@georgetown.edu}

\end{document}